# The InterHourly-Variability (*IHV*) Index of Geomagnetic Activity and its Use in Deriving the Long-term Variation of Solar Wind Speed

Leif Svalgaard

ETK, Inc., Houston, Texas, USA

Edward W. Cliver

Space Vehicles Directorate, Air Force Research Laboratory, Hanscom Air Force Base, Bedford, Massachusetts, USA

**Abstract.**

We describe the detailed derivation of the InterHourly Variability (*IHV*) index of geomagnetic activity. The *IHV*-index for a given geomagnetic element is mechanically derived from hourly values or means as the sum of the unsigned differences between adjacent hours over a seven-hour interval centered on local midnight. The index is derived separately for stations in both hemispheres within six longitude sectors spanning the Earth using only local night hours. It is intended as a long-term index and available data allows derivation of the index back well into the 19$^{th}$ century. On a time scale of a 27-day Bartels rotation, *IHV* averages for stations with corrected geomagnetic latitude less than 55° are strongly correlated with midlatitude range indices ($R^2$ =0.96 for the *am*-index since 1959; $R^2$ =0.95 for the *aa*-index since 1980). Assuming a constant calibration of the *aa*-index we find that observed values of *aa* before the year 1957 are 2.9 nT too small compared to values calculated from *IHV* using the regression constants based on 1980-2004. We interpret this discrepancy as an indication that the calibration of the *aa*-

22  index is in error before 1957. There is no systematic discrepancy between observed and

23  similarly calculated *ap*-values back to 1932. Bartels rotation averages of *IHV* are also

24  strongly correlated with solar wind parameters ($R^2 = 0.79$ with $BV^2$). On a time scale of a

25  year combining the *IHV*-index (giving $BV^2$ with $R^2 = 0.93$) and the recently-developed

26  Inter-Diurnal Variability (*IDV*) index (giving interplanetary magnetic field magnitude, *B*,

27  with $R^2 = 0.74$) allows determination of solar wind speed, *V*, from 1890-present. Over the

28  ~120-year series, the yearly mean solar wind speed varied from a low (inferred) of 303

29  km/s in 1902 to a high (observed) value of 545 km/s in 2003. The calculated yearly

30  values of the product *BV* using *B* and *V* separately derived from *IDV* and *IHV* agree

31  quantitatively with (completely independent) *BV* values derived from the amplitude of

32  the diurnal variation of the horizontal component in the polar caps since 1926 (and

33  sporadically further back).

34  **1. Introduction**

35  Modern geomagnetic indices aim at becoming proxies for solar wind parameters and to

36  be useful in studying the variation with time of the solar wind and, ultimately, the Sun.

37  While direct and systematic measurements of the solar wind extend a little more than

38  forty years, we have a geomagnetic record more than four times that long. In this paper,

39  we develop a new geomagnetic index, the InterHourly Variability or *IHV*-index, that

40  enables us to bring this extended record to bear on the question of the long-term variation

41  of the solar wind, a topic of increasing interest with impact on a range of solar-

42  heliospheric physics including the solar dynamo, climate change, and cosmic ray

43  modulation (*e.g. Fisk and Schwadron* [2001]*, Cliver et al.* [1998b], *Caballero-Lopez et*

44  *al.* [2004]).

## 1.1. Geomagnetic Indices Bear Witness to Solar Conditions

It was realized long ago [*e.g. Bartels*, 1940] that solar electromagnetic radiation (primarily Far UltraViolet, FUV) and solar "corpuscular" radiation (what we today call the "solar wind") give rise to conditions promoting different classes of fluctuations of the geomagnetic field. FUV radiation creates and maintains the lower ionospheric layers. Solar tidal motions of the ionosphere and thermally driven ionospheric winds produce a regular daily $S_R$ variation by dynamo action. The irregular geomagnetic variations are described or measured by geomagnetic *indices* that codify and compress the extraordinary complexity of the variations of the geomagnetic field. Modern geomagnetic indices attempt to remove the $S_R$ variation in order to isolate the irregular part ascribed to activity induced by the solar wind [*Mayaud*, 1980]. If this is possible, the index becomes a proxy for solar properties and can be used to study variation with time of the solar wind and the Sun.

## 1.2. *IHV*: A Mechanically Derived Long-Term Geomagnetic Index

Derivation of a geomagnetic range index [*Bartels et al.*, 1939; *Mayaud*, 1967] involves both the ability of the observer to correctly identify the variations not caused by the solar wind and the availability of appropriately intercalibrated conversion tables. Well-trained observers can obtain a remarkable consistency in scaling *K*-indices. But observers, stations, and instruments change over time, new conversion tables have to be drawn up and intercalibrated with the previous tables, and the station network thins when we go back in time. The entire process cannot easily or mechanically be duplicated and the quality and stability of calibration of the index values are difficult and labor-intensive to gauge. In the present paper we shall show that the long time-series of hourly values of the

68  magnetic components from observatories (some extending back into the 1830s) provide a
69  basis for constructing a new index, the InterHourly-Variability (*IHV*) index, measuring
70  solar wind related activity, without visual inspection of the original magnetograms (many
71  of which may no longer exist or - for eye-readings - may never have existed), using a
72  mechanical and readily duplicated derivation process.

73  **1.3. The Fundamental Difference Between *IHV* and Other Indices**

74  In deriving the *IHV* index, we follow the suggestion by *Mayaud* [1980] to exclude
75  daytime hours in order to eliminate the influence of the regular variation. Because we are
76  constructing an index for long-term trends, we can largely bypass the problem caused by
77  the solar FUV influence by only using data from the night-hours. During the seven hours
78  around local midnight, the influence of the $S_R$ variation is small and geomagnetic activity
79  is relatively largest. By only using a quarter of each local day we get an index that for a
80  given station is a statistical sample of the global activity. The sample is biased by any
81  UT-variations of activity, but this can be corrected for in a straightforward manner as
82  described below. By combining many stations distributed in longitude in both
83  hemispheres we aim to construct an index that on a 27-day rotation basis reproduces the
84  *am*-index [*Mayaud*, 1967] basically covering the entire geomagnetic record since regular
85  observations began. This procedure is validated *a posteriori* by the very close correlation
86  between our new index and the high-quality *am*-index. While the *am*-index is only
87  available since 1959, we obtain *IHV* in the present study back to 1883. Records exist, not
88  yet in digitized electronic format and therefore not yet incorporated into the index, that
89  should make it possible to extend the *IHV* continuously back to 1830s.

## 1.4. The Use of *IHV* to Derive the Solar Wind Speed

A close relationship between solar wind parameters and the *am*-index has been demonstrated by many workers. It is well-established [*e.g. Svalgaard*, 1977; *Feynman and Crooker*, 1978; *Murayama*, 1982; and *Lundstedt*, 1984] that geomagnetic range indices (such as *am*) are robustly correlated with the product $BV^2$, where *B* is the magnitude of the interplanetary magnetic field impinging on the Earth with solar wind speed *V*. We can therefore determine the product $BV^2$ from geomagnetic indices, including *IHV*. To make further progress we must *separate* the influence of the two parameters *B* and *V*. That, then, is the basic problem to be solved, as was realized by *Feynman and Crooker* [1978]. *Lockwood, Stamper, and Wild* [1999] attempted to determine *V* from an *ad hoc* formula involving the *aa*-index and the Sargent recurrence index [*Sargent*, 1986]. Although this approach worked well for the initial dataset employed, it failed when new data became available [*Svalgaard and Cliver*, 2006]. In the present study, we will use *B* derived directly from our Inter-Diurnal Variability index (*IDV*; *Svalgaard and Cliver*, 2005) to obtain *V* from $BV^2$ (*IHV*). We also use the product *BV* derived from polar cap diurnal variations [*LeSager and Svalgaard*, 2004] as a divisor for $BV^2$ to obtain an independent V-series with which to check that obtained by our primary method.

## 1.5. Why Not Use the Long-term *aa*-Index?

*Mayaud* [1972, 1973, 1980] established the *aa*-index as the standard measure of long-term geomagnetic variability. This index, based on observations from two nearly antipodal mid-latitude stations, one each in Australia and England, extends from 1868-present. The *aa*-index, that has an extension to 1844 [*Nevanlinna*, 2004], has been used in a variety of studies of long-term solar and solar-terrestrial variability, in particular, first by *Svalgaard* [1977] and then perhaps most notably by *Lockwood, Stamper, and Wild* [1999] to infer a more than doubling of the solar coronal magnetic field during the 20$^{th}$ century; claims that we cannot substantiate. These studies and numerous others (*e.g., Feynman and Crooker*, 1978; *Cliver et al.* [1998a]) assume a correct calibration of *aa* over time. Because of the growing use of the *aa*-index for our understanding of long-term solar behavior it is important to verify the long-term stability of its calibration. The *IHV*-index is based on many more stations than the *aa*-index and permits comparisons between several stations over extended periods of time rather than just between the single pair of stations over only two years that Mayaud used to calibrate *aa* after station changes. We confirm in the present paper several recent independent findings (*Jarvis* [2005], *Lockwood et al.* [2006b], and *Mursula and Martini* [2006], following the preliminary work in *Svalgaard et al.* [2003, 2004]) that the *aa*-index does not have stable calibration, is in need of revision, and is therefore, in its present version, not suitable for deriving quantitative information about long-term changes in the solar wind or the Sun.

**1.6. Roadmap**

After providing a detailed derivation of the *IHV*-index in sections 2-4 (with corresponding technical aspects discussed in Appendices A and B), we compare the *IHV* with the mid-latitude range indices, *am*, *ap*, and *aa*, in section 5. In section 6 we use the *IHV*-index to derive solar wind speed from 1890-present. We then substantiate this result by comparison with inferred solar wind parameters based on the polar cap potential back to 1926 and sporadically to earlier times.

## 2. Definition of the *IHV*-index

### 2.1. Historical Background

The *IHV*-index springs from the same source as the classical *u*-measure [*Bartels*, 1932], building on concepts by *Broun* [1861] and *Moos* [1910] who defined the interdiurnal variability *U* of the horizontal component at a given station as the difference between the mean values for that day and for the preceding day taken without regard to sign. The δ-index defined by *Chernosky* [1960] was a generalization of this inter-interval variability index, where δ could be taken as any value, not just one day or one hour. Both the *u*-measure and the δ-index suffer from contamination from $S_R$. *Chernosky* [1983] attempted to eliminate $S_R$ by computing the unsigned difference between corresponding three-hourly means on *successive* days, but was only partly successful, because $S_R$ itself varies from day to day. Our solution is more radical as our aim is somewhat lower. We do not attempt to construct an index value for every hour or three hours or even a day, but are content with a statistical sample based on the ¼ of the data when $S_R$ is not present. Such a sample, based on an can be expected to provide a reliable estimate of the average level of activity for intervals of weeks or longer, and will be almost free from contamination by $S_R$.

### 2.2. The Inter-Hour Variability

The *IHV* index is defined as the sum of the differences, without regard to the sign, of hourly means (or values) for a geomagnetic component from one hour to the next over the seven-hour interval around local midnight where the $S_R$ variation is absent:

$$IHV^{H} (nT) = \sum_{h=h1}^{h=h1+5} \text{abs}\, (H_h - H_{h+1}) \qquad (1)$$

158  where *h*1 is the starting UT-hour (0 to 23) of the interval. The hour *h* should be counted

159  modulo 24 to wrap around to the following day, if needed. $H_h$ is the hourly mean value

160  for the $h^{th}$ hour. If any of the hours in the interval does not have data, the *IHV* value is not

161  calculated for that day. The *IHV*-index can be defined for any geomagnetic component

162  (H, D, Z, X, Y, I, F) which may be denoted in an appropriate way, *e.g.* $IHV^H$ for *IHV*

163  derived from the H-component, which is what we concentrate on in the present paper. We

164  refer to *Jonkers et al.* [2003] for definition of geomagnetic components, their

165  relationships, and historical details of their measurement. Components that are expressed

166  as angles must be converted to force units (*e.g.* D (nT) = H (nT) · D (tenth of arc

167  minutes)/34377). The *IHV*-index must be calculated from data values given or properly

168  rounded [*Ellis*, 1900] to the nearest 1 nT. The *IHV*-index for a given station is computed

169  as one value for the UT-day that contains roughly local midnight (see Section 3.1), but is

170  not a *daily* index, as we only sample part of the day. An average over an interval of many

171  day values (*e.g.* over a month or a 27-day Bartels rotation) is expected to approximate the

172  average activity over the interval because geomagnetic activity has a high degree of

173  conservation [*Chapman and Bartels*, 1940, p. 585].

174  **2.3. Demonstration of *IHV* Determination for a Single Station**

175  The geomagnetic observatory at Fredericksburg, Virginia (FRD) has been in operation

176  since 1956 (with a brief data availability gap 1981-1983). This observatory is located

177  close to the ideal geomagnetic latitude of 50° for discerning the class of activity used in

178  derivation of the *am*-index (and is, in fact, one of the stations contributing to the *am*-

179  index - and to the *ap*-index, as well).

Figure 1 shows the variation of the three components (H, D, and Z) for several days in May 1999. The $S_R$ variation is clearly seen, including its day-to-day variability. An "effective" noon can be defined as the time where the H-component has its maximum excursion. This is also the time when the excursion in the D-component changes sign. An effective midnight is then 12 hours away. It is evident for this station, that 00-06 UT is a suitable interval for calculation of the *IHV*-value as the $S_R$ variation is minimal during this time. We have preliminarily chosen this interval (although in the end we use an interval starting one hour later for FRD - see section 3.1) because it just contains the first two 3-hour intervals of the UT-day. We can thus readily compare our *IHV*-values with the corresponding *am*-values. We can compute the average *am*-value over the six-hour interval for the day for direct comparison with the *IHV*-value for the day. We denote this average value by *Am2* to distinguish it from the daily average (not to be confused with the average over two days).

Figure 2 shows how well our new index compares to the *Am2*-index on a time scale of a month over the (arbitrary) interval 1970-1976. The *IHV*-index has a high correlation with the *Am2*-index (coefficient of determination $R^2 = 0.88$). This close agreement between the two indices even on a time scale as short as a month is the primary argument for the validity of our approach. For a time scale of one day, $R^2$ is still as high as 0.74. The finding that a single station for a limited local time range suffices to fairly characterize global geomagnetic activity during that time is well known and is also the underlying rationale behind the *aa*-index.

## 3. Selection of Stations, Local Time Interval, Data, and Sectors

Table 1 shows coordinates and dates of availability for the stations selected for this study. We have shown that a useful index can be derived from even a single station regardless of its location - as long as it is well equatorward of the auroral zones (section 3.3). The stations shown in Table 1 have been selected based on the availability of electronically readable hourly data from the World Data Centers for Geomagnetism, the INTERMAGNET program, and other sources. More data exist (in yearbooks and observatory reports), but are typically not yet available in electronic form. In this section we detail the criteria for choosing the local time interval, latitude regions appropriate for *IHV*, and longitudinal sectors, as well as the method used to compensate for the UT variation of geomagnetic activity (undesirable in an index aimed at being a proxy for solar conditions).

### 3.1. Selection of Interval During the Night

Because the regular diurnal variation is controlled mainly by the sun's zenith angle, we use the six-hour interval centered on ordinary geographical midnight, rather than using geomagnetic midnight. The difference is only important at high latitudes where, as we shall see, *IHV*, like the *K*-index, should not be used anyway in the meaning of a subauroral zone index. The data available to us are described in the WDC-format as "hourly means centered on the Universal Time half-hours". In reality, the 24 values per day may refer to hourly means centered on the half-hour, or on the whole-hour, or to instantaneous values measured on the whole-hour, or the half-hour. Moreover, "hour" may refer to UT, local standard time, true local time, "Astronomical" time or even

Göttingen time. These variations (at times unknown and inferable only from the data themselves by determining the time of "effective" noon from the diurnal variation by visual inspection) make it difficult to devise a hard rule for assigning the time of the interval to use. The solution we have chosen is to manually assign a number of "hourly" data values to *skip* at the start of a time series for a given station (purportedly starting as an hourly mean centered on $00^h30^m$ UT and labeled as hour 00) to align the series with "the six-hour interval around local geographic midnight" determined by visual inspection of the data. This number is usually within one hour of the time, *h*, calculated from $h = $ (rounded (24 - *longitude*/15) - 3) modulo 24, and is given in Table 1 as well. The values of *IHV* are not very sensitive to small variations of the exact starting time of the interval as can be seen from Figure 3. Using a result from section 6.2, namely that *IHV* is strongly correlated with the quantity $BV^2$ calculated from observed values of the IMF magnitude, *B*, and the solar wind speed, *V*, we show the coefficient of determination, $R^2$, as a function of the number of hours to skip for the Kakioka observatory (KAK) to obtain optimum correlation. Note the very broad maximum showing the insensitivity of the precise number of hours to skip, as long as the value chosen is within the broad range of high correlation.

After skipping to the beginning of the six-hour interval, the following seven hours are then used to calculate six unsigned successive differences between the hourly data; we referred to this as the "six-hour" interval. Their sum is the "raw" *IHV* value for the UT-day containing the forth hourly data point. If any of the seven data values are designated as "missing", *IHV* is considered missing for that day. Finally we skip over the 24 - 7 = 17 following hourly data points, positioning to the next six-hour interval and repeat the procedure until the end of the dataset. We assume that missing data are marked by a special value rather than by the absence of data entries.

**3.2. Removing Universal Time Variation**

*Svalgaard* [1977], *Svalgaard et al.* [2002] and references therein, and *O'Brien and McPherron* [2002] show that the equinoctial mechanism component of the semiannual variation of geomagnetic activity (*e.g.* as expressed by the *am*-index) can be closely described as a modulation of existing activity by a function of the form

$$(1 + 3 \cos^2\Psi)^{-2/3} = S(\Psi) \qquad (2)$$

where $\Psi$ is the angle between the solar wind flow direction and the Earth's magnetic dipole axis. The function $S(\Psi)$ varies both with the day of the year and with Universal Time (UT), and also very slowly due to secular variation of the geomagnetic field and (slower yet) the Earth's orbital elements; the total secular variation of $S$ since 1800 does not exceed 1%, thus is not yet measurable. Figure 4 shows the variation of the *S*-function (bottom panel) and of the "raw" *IHV* (top panel) with month of year and Universal Time calculated for *all* the stations in Table 1, for *all* data available for each station. The *IHV* values for a given station were assigned to the Universal Time of local midnight for that station. All values were divided by the average values for each station to make them comparable. The diminution of activity at the solstices when the geomagnetic pole on the sunward side comes closest to the subsolar point is clearly seen. Indeed, the variations of *S* and of *IHV* are *quantitatively* very similar, suggesting that we may remove the equinoctial mechanism part of the UT-variation of *IHV* by the simple expedient of dividing by *S* for the average UT-time for each single value of the differences defining *IHV* for each station and for each day. Removing this UT-variation minimizes the ill effect of (the inevitable) uneven station distribution. In the polar caps the UT-variation is small and is swamped by the large seasonal variation caused by the variation of ionospheric conductivity, being much higher during local summer. As we shall see, the *IHV*-index should not be used for stations within the polar cap, so removal of the UT-variation for such stations is moot.

Several workers (*e.g. Svalgaard et al.* [2002]; *Cliver et al.* [2000]; and *Crooker and Siscoe* [1986]) have suggested that typically some 25% of the semiannual variation is caused by other mechanisms than the equinoctial effect, possibly related to external causes (variations of solar wind properties with heliographic latitude (axial mechanisms) or to the angle between the magnetic axes of the Sun and the Earth (Russell-McPherron effect)). On rare occasions (*e.g.* during 1954), these effects can be large or even dominant [*Cliver et al.*, 2004]. We leave those external variations in the index.

**3.3. Latitudinal Variation of the *IHV*-index**

To get maximal spatial coverage we computed $IHV^H$ freed from the $\Psi$-related variation as described in section 3.2 for *all* 128 stations which had submitted data to the WDCs for the interval 1996-2003, covering most of a solar cycle. Since there were missing data at some stations at different times, the following procedure was used to make the data comparable. The stations were divided into six longitudinal sectors centered on (East) longitudes 15º, 75º, 130º, 200º, 240º and 290º. <*IHV*> was computed for the H-component for each station as a mean over 27-day Bartels rotations for which 20 or more days had data. Within each longitude sector a reference station was chosen that had good data coverage. The Bartels rotation averages for all stations within a sector were then

291 normalized to the reference station (by dividing by <*IHV*> for the reference station). The
292 overall averages <*IHV*> covering the full eight-year interval for the reference stations
293 were then themselves normalized to the overall <*IHV*> for NGK, and each station finally
294 normalized to that same standard by multiplying by the normalized reference averages.
295 The meaningful application of such a procedure relies on the assumption that *IHV*-values
296 at different stations are related through a simple constant of proportionality. This
297 assumption is found to hold to a degree of accuracy, arguably good enough for the
298 purpose of establishing the variation of *IHV* with latitude.

299 Plotting the normalized <*IHV*> for all stations against their corrected geomagnetic
300 latitude we obtain Figure 5. Other measures of latitude (geographic, simple geomagnetic,
301 dip) result in larger scatter. It is evident that the latitudinal variation is weak below
302 corrected geomagnetic latitude 55º, but that *IHV* increases sharply (by an order of
303 magnitude) on the poleward side of that latitude. Distance from the auroral zone seems to
304 be critical and we stipulate that *IHV* should only be used with its ordinary meaning for
305 stations that are equatorward of 55º. A similar stipulation holds for the well-known *K*-
306 index. In fact, the strong dependence on latitude above 55º might possibly afford a means
307 to monitor the long-term variation of the position of the auroral zone.

308 Also shown in Figure 5 is a fit to the data points by a simple *ad hoc* model consisting of
309 the sum of four Gaussian functions, *viz*.

310 $$IHV(\varphi) = \sum_k [a_k \exp\{-b_k [\text{abs}(\sin(\varphi)) - \sin(\varphi_k)]^2\}] \qquad (3)$$
311

312 where $a_k$ (k = 1,.., 4) define the scale (relative to a station at φ = 48º (*i.e.* NGK) with
313 <*IHV*> = 34.33 for 1996-2003), $b_k$ the width, and $φ_k$ the position of the peaks. These
314 parameters are given in inline Table 4.

315 [Table 4]

| K | $a_k$ | $b_k$ | $φ_k$º |
|---|-------|-------|--------|
| 1 | 0.218 | 10    | 0      |
| 2 | 0.728 | 0.04  | 23.5   |
| 3 | 9.700 | 435   | 68     |
| 4 | 0.408 | 6     | 90     |

316

317 The values of $φ_k$ were prescribed except for k = 3, which represents a least squares fit.
318 The model is descriptive only and does not pretend much physical content with the
319 exception of the identification of the auroral zone peak. As illustrative of the sensitivity
320 of *IHV* to the location of the auroral zone we note that a change of three degrees of either
321 φ or $φ_3$ entails a change of *IHV* by a factor of two for φ near 60º.

322 **3.4. Cautionary Notes and Technical Details**

323 In the course of developing the *IHV* index, we encountered a number of technical issues
324 and concerns related to the data used. These are covered in detail in Appendix A.
325 Although relegated to an Appendix, these issues are of utmost importance and must be
326 dealt with properly. They include:

327     (1) Non-removal of "Residual" Regular Variation;
328     (2) Use of hourly values instead of hourly means in older data;

(3) Effect of extreme activity;

(4) Station specific artifacts in archival data;

(5) The effect of secularly changing geomagnetic dipole position and strength;

(6) Missing data during severe storms;

(7) Possible seasonal bias.

**3.5. Longitude Sectors, Hemispheric Series, and Equatorial Series**

Although the geographic distribution of geomagnetic observatories suffers from the usual deficiencies of coverage over oceans and in the Southern Hemisphere, we have found it possible to construct long time series of *IHV* for the six longitude sectors described in section 3.3. In each sector we define *IHV* series separately for the two hemispheres yielding a total of 12 independent *IHV* series with various degrees of time coverage. In addition, we construct a series using stations close to the geomagnetic equator (within 9º). For the purpose of comparing and normalizing *IHV* for different stations, the series are calculated as averages over Bartels rotations (with at least 20 days of data). Figure S1 through S12 in the Electronic Supplement shows the *IHV*-series for each station within each longitude sector.

The longest of all series is constructed from the German stations Potsdam (POT: 1890-1907), Seddin (SED: 1908-1931), and Niemegk (NGK: 1932-present), supplemented with data from Wilhelmhafen (WLH: 1883-1895). This series of superb observations serves as a 'reference' series, all other series being calibrated and normalized to the WLH-POT-SED-NGK series. The normalization is necessary because of different underground electrical conductivity or seawater 'coast' effects. These effects often exceed the small effect of differing latitudes so we have adopted the practice of not having a separate latitude effect on top of the overall normalization factors, avoiding Mayaud's [1973, p.7] mistake of correcting *twice* for latitude.

**4. Calibration of Longitude Sectors and a Global Index**

**4.1. Calibration of Single Sectors: the Principle of Overlap**

Within each sector we select stations such that there is maximal *overlap* in time. Two stations that have data that does not overlap can be compared if a third station, or better, as is often the case, several stations, overlap with the stations to be compared such as to form one or more "bridges". For overlapping data we perform a linear regression analysis. Figure 6 shows a typical example. Rare outliers (one is marked by the large open circle) are identified and are not included in calculating the linear slope (or "scale") between the two stations. The "offset" is usually small to insignificant, so we force the offset to zero. The square of the linear correlation coefficient ($R^2$) indicates how good the fit is. Because both of the datasets that are being correlated have errors or variance not accounted for by the correlation, the "usual" method (that minimizes the sum of the squares of the ordinate deviations) is not applicable. The "proper" way of dealing with this problem has a 100-year literature [*e.g. Parish*, 1989]. We use a method equivalent to minimizing the sum of the squares of the perpendicular distances along both axes to the best fit-line, rather than just the vertical distances. The correlations are usually good enough that the problems associated with "constrained" regression are not a concern (*e.g.* in calculating $R^2$). The data has heteroscedacity (larger values scatter more than smaller values), but this is mostly offset by larger values being progressively rarer. In reporting slopes we shall quote four decimals throughout, being sufficiently (even overly) precise that rounding errors are avoided. In Appendix B we detail the construction and calibration of each sector.

An overriding principle has been to avoid hidden empirical adjustments, and in cases where normalizations were used, to make these reversible and transparent. Anyone should be able to recalculate the index from public archival data. As more data become available, index values may change slightly. There will thus not be a definitive "official" version of a "global" *IHV*-index. In this we recognize the importance of future recalibration (by anybody) based on improved data and understanding. We propose that a given version of *IHV* be annotated with the year of its release, *e.g. IHV2007* for the version derived in the present paper.

**4.2. The Composite *IHV* Series**

A composite *IHV* series can be formed by first adding the average *IHV* values for all of the longitudinal sectors that have data for each Bartels rotation. To compensate for the fact that data may be missing for certain sectors from time to time, the summed *IHV* for each rotation is divided by the number of sectors that contributed data for that rotation, effectively calculating the mean with each sector having the same weight. The composite *IHV* now becomes a true daily index: successive longitude sectors contributing their six-hour slices as the Earth turns. The variation from sector to sector is simply the variation of geomagnetic activity with time. Because geomagnetic activity has a high degree of conservation, the *IHV*-index for one sector is strongly correlated with the *IHV*-index for the following sector(s). It is this property that makes it possible to normalize all sectors to the Northern Hemisphere part of the sector centered on longitude 15º (called IHV15N, effectively to NGK) with the scaling factors given in Table 5. Since all sectors are ultimately reduced to NGK, any *intrinsic* variation (apart from that related to the dipole tilt, see section 3.2) of geomagnetic activity with longitude or with hemisphere is not reflected in the *IHV* series.

[Table 5]

| From | IHV15N | $R^2$ | Time |
|---|---|---|---|
| IHV15N | 1.0000 | 1.00 | 1890-2006 |
| IHV15S | 1.0022 | 0.84 | 1932-2004 |
| IHV75N | 1.2335 | 0.77 | 1925-2004 |
| IHV75S | 1.2926 | 0.77 | 1957-2004 |
| IHV130N | 1.4563 | 0.63 | 1913-2006 |
| IHV130S | 1.3446 | 0.60 | 1919-2004 |
| IHV200N | 1.5445 | 0.59 | 1902-2004 |
| IHV200S | 1.5146 | 0.43 | 1922-2004 |
| IHV240N | 1.1188 | 0.60 | 1910-2004 |
| IHV240S | 1.3333 | N/A | 1964-1964 |
| IHV290N | 1.1969 | 0.73 | 1903-2004 |
| IHV290S | 1.2202 | 0.60 | 1915-2005 |

The top panel of Figure 7 shows a portion of the individual data series that went into the composite series. Northern sectors are shown in black while southern sectors are shown in red. A rotation by rotation plot of the full series is shown in Figure S14 of the Electronic Supplement. The middle panel of Figure 7 shows the full series overlain by its 13-rotation running mean. The lower panel of Figure 7 shows 13-rotation running means of the composite *IHV* (blue) and *IHV* derived from Equatorial stations (red). The entire composite dataset is given in Table T1 of the electronic supplement.

**4.3. No Seasonal Variation of the *IHV*-index**

We can also construct composite *IHV*-series for each hemisphere separately. Because of the normalization of each sector to IHV15N there will be no general difference in activity level between hemispheres, but it is of interest to investigate the variation of the index with time (annual phase) of year since one of the rationales for constructing indices based on the mean of antipodal observations (such as *aa*) is to cancel out any variation tied to the seasons. We show in Appendix A.7 that there does not seem to be any significant seasonal variation, *i.e.* related to summer/winter differences. This is important for construction of a global index (usable for inferring solar properties) from *IHV* when we only have data for one hemisphere (at the time of writing there is very little Southern Hemisphere data before 1915, see Table 5). With no seasonal (*i.e.* summer/winter) variation such missing data is not going to skew the result in any systematic or appreciable way.

**5. Comparison with Geomagnetic Range Indices**

In this section we compare the composite *IHV* series with the family of range indices comprising the *am*, *ap*, and *aa*-indices with emphasis on long-term stability. The composite *IHV*-index is derived from many longitudinal sectors in each hemisphere comprising many stations with much overlap of data. The excellent agreement between all these independent series is interpreted as confirmation of the long-term stability of the composite *IHV*-index. The Electronic Supplement contains detailed, rotation by rotation comparison plots (Figures S15 and S16).

**5.1. Comparison with the *am* Index**

Figure 8 shows the relationship between rotation means of (composite) *IHV* and the *am*-index. As with *IHV*, we have removed the dipole tilt-related variations by dividing *am* by the *S*-function (eq.(2)). The clearly non-linear relation can be expressed as:

$$Am = 0.2375 \, IHV^{1.2892}, \qquad (R^2 = 0.96) \qquad (4)$$

$R^2$ is calculated from the linear relationship between the logarithms. The excellent (near perfect) fit means that we can use *IHV* as a proxy for *Am* (and similar range indices). Figure 9 shows the *Am* proxy calculated from *IHV* using eq.(4) *in extenso* for every Bartels rotation where we have data.. The *am*-index is probably the best range index available at this time, based as it is on a satisfactory station distribution. We urge the reader to study this compelling Figure.

**5.2. Comparison with the *ap* Index**

Figure 10 shows the relationship between rotation means of (composite) *IHV* and the *ap*-index. As with *IHV*, we have removed the dipole tilt-related annual and UT-variations by dividing *ap* by the *S*-function (eq.(2)). Because the UT variation was deliberately sought eliminated when the *Kp* index tables were drawn up [*Bartels et al.*, 1939], one might try to evaluate the *S*-function with UT set to a constant value (say 12) for every 3-hour interval during the day. This, however, results in a markedly lower correlation ($R^2 = 0.85$), so we resort to using the actual UT. The clearly non-linear relation can be expressed as:

$$Ap = 0.0549 \, IHV^{1.5596}, \qquad (R^2 = 0.91) \qquad (5)$$

450 The correlation is significantly worse than for the *am*-index, reflecting the high quality of
451 the *am*-index. Figure 11 shows the *Ap* proxy calculated from *IHV* using eq.(5). Inspection
452 of the Figure shows that there is no systematic drift between the observed and calculated
453 values of *Ap* over the entire interval 1932-2004. For no 10-year interval does the absolute
454 difference between the observed and calculated averages of *Ap* exceed 5%. This means
455 that we can use *IHV* as a stable proxy for *Ap*. There are, however, from time to time for
456 intervals of a few years, systematic differences showing that *Ap* is not quite homogenous.

457 **5.3. Comparison with the *aa* Index**

458 Figure 12 shows the relationship between rotation means of (composite) *IHV* and the *aa*-
459 index for the time interval (1980-2004) since the latest (published) change of the *aa*
460 calibration table values. As with *IHV*, we have removed the dipole tilt-related annual and
461 UT-variations by dividing *aa* by the *S*-function (eq.(2)). The clearly non-linear relation
462 can be expressed as:

463 $$Aa = 0.3600 \, IHV^{1.1856}, \qquad (R^2 = 0.95) \qquad (6)$$

464 The excellent fit (almost as good as for *Am*) means that we should be able to use *IHV* as a
465 proxy for *Aa* as well. Figure 13 shows the *Aa* proxy calculated from *IHV* using eq.(6).

466 As Figure 13 shows, the observed values of *Aa* match the calculated proxy values very
467 well back in time until about the beginning of 1957. Before that time, observed *Aa* is
468 consistently smaller than calculated *Aa*. Figure 14 shows the difference in a more
469 compact format. The jump at 1957.0 is 2.9 nT or 12% of the average value of *Aa*. At
470 times where *Aa* is much smaller, such as at the beginning of the 20$^{th}$ century, the
471 percentage discrepancy is much larger (~40%). We interpret the difference to be an indication that the calibration of *aa* as measured by rotational means before 1957 is in error by this amount. A similar conclusion was already reached by *Svalgaard et al.* [2003, 2004], *Jarvis* [2005], *Lockwood et al.* [2006b], and *Mursula and Martini* [2006]. A critical re-examination and recalibration of the *aa*-index will be covered in a subsequent paper in this series. It would seem that *IHV* could serve as a useful tool for checking the stability of geomagnetic indices, both for past values and for ongoing quality control. That such ongoing control is needed should be clear from the account by *Lincoln* [1977] in *van Sabben*, [1977].

## 6. Comparison with External Solar Wind Drivers

The earliest solar wind data showed strikingly that geomagnetic activity depends strongly on solar wind speed [*Snyder et al.*, 1963]. It is well-established [*e.g. Svalgaard*, 1977, *Murayama*, 1982, and *Lundstedt*, 1984] that geomagnetic range indices (such as *am*) are robustly correlated with $P = q\,BV^2$, where $B$ is the magnitude of the interplanetary magnetic field impinging on the Earth with solar wind speed $V$. The geometric factor $q$ parameterizes the effect of magnetic merging depending on the angle (and of its variability over the 3-hour interval) between the interplanetary and terrestrial magnetic fields. We shall initially assume that the average $q$ over a rotation does not vary from one Bartels rotation to the next. We also ignore a very weak dependence of solar wind density $n$, in the form $n^{1/3}$ [*Svalgaard*, 1977]. We show in section 6.2 that the *IHV*-index is a sensitive indicator of $P$, responding directly and simply to this external solar wind driver.

## 6.1. Solar Wind Data

*In situ* near-Earth solar wind data is available from the OMNIWEB website at `http://www.omniweb.org`. We utilize Bartels rotation averages of *B* and *V*. First, daily averages are calculated from hourly averages. If there is any data at all for a given day, its daily average goes into the rotation average. If the rotation average is based on less than 20 days of daily averages, the rotation is not used. For some years (especially during the 1980s) there were significant amounts of data missing. Over a time scale of 27 days and longer there is little difference between $<B><V>^2$ and $<BV^2>$. We use the separable version, $<B><V>^2$, because our ultimate goal is to calculate $<V>$. We shall often use the abbreviation $V_o$ for the quantity $V/(100\text{ km/s})$. When we calculate regression lines involving interplanetary parameters we have treated those as dependent variables with errors (*e.g.* caused by missing data) assuming *IHV* to be 'error free'. This is guided by our wish to see how well we can estimate $BV_o^2$ from *IHV* and not how well we can calculate *IHV* from $BV_o^2$ (see the exchange in *Lockwood et al.* [2006a] and *Svalgaard and Cliver* [2006]). Available interplanetary data are for several years only poorly representative of true solar wind conditions because of significant amounts of missing data (approaching 60% or more during 1983-1994).

## 6.2. *IHV* Dependence on $BV^2$ (Rotation Time Scale)

Figure 15 shows the correlation between the composite *IHV* series and $BV_o^2$. Assuming the simple linear form suggested by the Figure, the relationship can be written:

$$BV_o^2 = (4.33\pm0.11)\,(IHV - 6.4\pm1.0), \qquad\qquad R^2 = 0.77 \qquad (7)$$

513   Figure 16 shows how well *IHV* reproduces $BV_o^2$. There is detailed agreement even on a
514   time scale as short as one rotation.

515   Figure 17 shows 13-rotation (~1 year) running means of calculated $BV_o^2$ and observed
516   values of $BV_o^2$ back to 1965. Close examination shows systematic disagreements
517   concentrated in certain years: 1975-1978 and 1995-1998. It is no coincidence that these
518   intervals are ~22 years apart. The Russell-McPherron effect [*Russell and McPherron*,
519   1973] gives rise to a semiannual variation of geomagnetic activity that usually is very
520   small, *except* when the Rosenberg-Coleman effect [*Rosenberg and Coleman*, 1968;
521   *Wilcox and Scherrer*, 1972] is pronounced. *Echer and Svalgaard* [2004] found that the
522   Rosenberg-Coleman effect tends to occur only near sunspot minimum and then for a few
523   years thereafter during the rising phase of the solar cycle as shown in the Figure by the
524   amplitude of the R-C effect determined by wavelet analysis. The R-M effect plus a
525   strong, short-lived R-C effect combined with a reversal of the large-scale solar polar
526   fields gives rise to a few years of enhanced geomagnetic activity every ~22 years
527   [*Chernosky*, 1966; *Russell and Mulligan*, 1995; *Cliver et al.* 1996] that will result in *IHV*
528   being larger than usual during such times. The two short periods of (minor)
529   disagreements between the observed and calculated values of $BV_o^2$ were just two such
530   times. Another one (and an extreme one at that) was the year 1954 [*Cliver et al.* 2004]
531   and some disagreements would be expected around 1934-1935, 1913-1915, 1890-1892,
532   *etc*.

533   If the few years of 22-year cycle 'contamination' are not included in the fit, the
534   relationship between $BV_o^2$ and *IHV* becomes

$$BV_o^2 = (4.25\pm0.11)\ (IHV - 5.0\pm0.9)\ , \qquad\qquad R^2 = 0.79 \qquad (8)$$

The smallest values of *IHV* averaged over a rotation in the ~120-year series are around 13 (only 5 values out of more than 1600 are smaller than 14). By way of illustration, *IHV* ~13 gives $BV_o^2$ ~34 which would be satisfied by $B = 4.5$ nT and $V = 275$ km/s.

The excellent agreement between observed and calculated values of $BV_o^2$ even before 1974 suggests that the interplanetary measurements are of high quality and that one cannot maintain that the accuracy of the IMF and solar wind data was low during the "baby" period of the space age (*e.g. Stoshkov and Pokrevsky* [2001]) as a reason for differences between inferred and observed parameters.

### 6.3. *IHV* Dependence on $BV^2$ (Yearly Time Scale)

We compute the yearly mean of *IHV* (or of $BV_o^2$) for a given year by averaging over Bartels rotations spanned by the year. Figure 18 shows the relationship between yearly means of $BV_o^2$ and *IHV*. Omitting the few years of 22-year enhancements, yields this regression equation for yearly means:

$$<B><V_o>^2 = (4.34\pm0.21)\ (<IHV> - 6.2\pm1.9)\ , \qquad\qquad R^2 = 0.93 \qquad (9)$$

We may note that all the regression equations (7) through (9) are identical within their statistical errors.

### 6.4. Determination of Solar Wind Speed

In *Svalgaard and Cliver* [2005] we showed how yearly averages of *B* could be determined from our *IDV*-index:

$$<B> = (3.04\pm0.37) + (0.361\pm0.035)\ <IDV>\ , \qquad\qquad R^2 = 0.74 \qquad (10)$$

Combining eqs.(9) and (10) yields

$$\langle V\rangle = 347 \text{ km/s } [(\langle IHV\rangle - 6.20)/(\langle IDV\rangle + 8.42)]^{1/2}, \tag{11}$$

allowing determination of V from IHV and IDV. Figure 19 shows B and $V_o$ since 1890. Over the ~120-year series, the solar wind speed varied from a low (inferred) value of 303 km/s in 1902 to a high (observed) value of 545 km/s in 2003. Table 2 gives the yearly average values of IHV, B and V calculated as the average of the rotations spanned by each year.

**6.5. Comparison with BV derived from Polar Cap Potential**

*Le Sager and Svalgaard* [2004] derived yearly averages of BV from a study of the amplitude of the diurnal variation of the geomagnetic elements recorded at the polar cap stations Thule and Godhavn. The horizontal component of the geomagnetic field measured within the polar regions has a particularly simple average diurnal variation: the end point of the component vector describes a circle with a diameter (the "range", E) of typically 100 nT. E is controlled by season (ionospheric conductance) and by the interplanetary electric field as measured by BV mapped down along field lines to the polar cap. Averaging over a full year eliminates the seasonal dependence on conductivity and yearly average ranges have a strong ($R^2 = 0.9$) linear relationship $BV = kE$ with BV calculated from B and V measured by spacecraft. This relationship holds for any station within the polar caps with only a slight variation of k. A small constant term is not statistically significant, so is not considered further. We have determined k for three polar cap stations: Thule (data back to 1932, k = 24.9 for V in km/s, B and E in nT) and Godhavn (back to 1926, k = 32.0) in the northern polar cap and Scott Base (back to 1957, k = 27.9) in the southern polar cap. Figure 20 compares the product BV calculated from B and V derived from the IDV and IHV indices and derived from the polar cap stations. We note substantial quantitative agreement between these completely independent determinations.

### 6.6. Comparison with Polar Cap Potential from 1902-1905

The noted Norwegian explorer Roald Amundsen wintered over with his ship "Gjøa" at "Gjøahavn" in Northern Canada close to the magnetic pole. Magnetic recordings began on November 1$^{st}$, 1903 and continued through May 1905 [*Steen et al.*, 1933]. The National (British) Antarctic Expedition of 1901-04 under Robert F. Scott operated magnetographs at the Winter Quarters (sometimes known as Discovery Bay or Hut) of the Expedition for nearly two full years (1902-03) [*Chree*, 1912]. The range of the diurnal variation has been determined for these two sets of observations and is shown in Figure 21. Magnetographs were in operation at Cape Evans, the base station of the British (Terra Nova) Antarctic Expedition during 1911 and 1912. Cape Evans and Winter Quarters are co-located with Scott Base. The range was also determined for this station. However, this value represents only a lower limit to the true range because disturbed days, where the trace was not complete, were excluded. These early determinations are also shown in Figure 20 using *k*-values derived from the modern data (using the *k*-value for Godhavn (Qeqertarsuaq) for Gjøahavn, as these two stations have nearly the same corrected geomagnetic latitude).

In order to compare with modern values we select years (for SBA) where the sunspot number $R_z$ was similar [*viz.* 14] to $R_z$ in 1903, namely 1965, 1975, 1976, 1985, 1986, 1995, and 1996. For GDH we select years where $R_z$ was similar [42] to $R_z$ in 1904 (and

601 on the ascending branch only), namely 1966, 1977, 1987, 1997, and 1998. Then we
602 derive *BV* for the modern years using observed *B* and *V*:

603 [Table 6]

| Station | IAGA | Year | $<R_z>$ | CGMlat | ΔY nT | ΔY' nT | BV | V | B | B IDV |
|---|---|---|---|---|---|---|---|---|---|---|
| Winter Quarters | HUT | 1903.0 | 14 | -81.2° | 83.9 | *78.4* | 2316 | *446* | *5.19* | 5.02 |
| Gjøahavn | GJO | 1904.5 | 42 | 79.3° | 78.1 | *71.1* | 2427 | *414* | *5.86* | 5.53 |
| Scott Base | SBA | >1964 | 14.4 | -79.9° | 84.4 | 84.4 | 2493 | 446 | 5.71 | 5.94 |
| Qeqertarsuaq | GDH | >1964 | 37.9 | 77.0° | 75.0 | 75.0 | 2560 | 414 | 6.17 | 6.16 |

604
605 The range of the polar cap variation is measured using the Y-component (in a local
606 coordinate system where the X-axis coincides with the average direction of the H-
607 component). Going back to ~1900 the main field increases 6.5% at SBA/HUT and 9% at
608 GDH/GJO. If we decrease the ionospheric conductivity by the same amounts, we
609 decrease Δ*Y* proportionally to Δ*Y'* as shown. Assuming *BV* scales with Δ*Y'*, we get for
610 HUT: *BV* = 78.4/84.4*2493 = 2316, and for GJO: *BV* = 71.1/75.0*2560 = 2427. If *V* in
611 1903 were the same as for the modern group of years (446 km/s) for SBA, we obtain *B* =
612 2316/446 = 5.19 nT. If *V* in 1904 were the same as for the modern years (414 km/s) for
613 GDH, we obtain *B* = 2427/414 = 5.86 nT. These values compare favorably with *B*
614 derived from the *IDV*-index. If the solar wind speed a hundred years ago were somewhat
615 lower (as we deduce in this paper), *B* would be correspondingly, if only slightly, higher.
616 We take this as evidence for *B* ~100 years ago not being lower than now by any
617 significant amount.
618

## 7. A Plea for Assistance with Early Data

Table 3 shows an incomplete compilation of early observatories that have data suitable for calculation of *IHV*. Because almost none of the pre-1920 data is available from the World Data Centers, the authors ask the geomagnetic community for help in collecting and preserving the large body of early data, if you have any access to or knowledge of the whereabouts of data from stations listed in the Table. There is a vast amount of $19^{th}$ and early $20^{th}$ century data in existence. An effort should be made to move all available data into electronic form for general use by all. Some examples: *Nevanlinna* [2003] reports a modern reduction of 10-minute observations from Helsinki 1844-1857 and with coarser resolution until 1912. *Moos* [1910] describes the superb Colaba observations going back to 1846. *Chapman's* [1957] analysis of the solar daily variation was based on "more than a million hourly values of the magnetic elements [from Greenwich]. This great series of observations was begun in 1838 by Airy". *IHV*-indices could possibly be constructed from nighttime subsets of these and other observations, once the data is put in digital form. There are early $19^{th}$ century data from Paris, Prague, Milan, Munich, and other places. It is quite possible that some of that data may be usable for derivation of approximate *IHV* indices.

## 8. Conclusion

In the present paper we have provided a detailed derivation of the *IHV*-index and used it, in conjunction with the newly-developed *IDV*-index [*Svalgaard and Cliver*, 2005] for a reconstruction of solar wind speed from 1890-present. In addition, comparison of the *IHV*-index with the widely used *aa*-index reveals calibration errors in *aa* prior to 1957 as suggested by *Svalgaard et al.* [2003, 2004] and substantiated by others (*Jarvis* [2005], *Lockwood et al.* [2006b], *Mursula and Martini* [2006]). The *IHV*-index will need to incorporate additional early data, both to corroborate data from ~1880-1920 and to extend the index back in time. Such work in is progress with preliminary results already back to 1844.

**Appendix A.**

**A.1. Non-removal of "Residual" Regular Variation**

*Mursula et al.* [2004] suggest that it is necessary to identify and remove any possible "residual" $S_R$ variation before calculating *IHV*. The main argument for this is that the diurnal variation of some of the geomagnetic elements does not show the characteristic flat "plateau" at high-latitude stations such as SIT and SOD during the night hours (themselves somewhat ill-defined at high latitudes). At such high latitudes the signatures of the eastward and westward auroral electrojets are clearly seen before and after midnight, respectively. We would ordinarily consider those signatures as part of "geomagnetic activity" so see no need to remove them (apart from not knowing how to do this in any non-objectionable way). The proper thing to do is simply not to treat *IHV* computed for high-latitude stations as comparable to *IHV* calculated from mid- and low-latitude stations. Figure 22 shows the local-time diurnal variation of the unsigned difference between the values of the horizontal component for one hour and the next for bands of corrected geomagnetic latitude intervals from the equator to the pole. Note the effect of ring current decay through the day at mid-and low-latitude stations. It is clear that stations above ~55º have a different activity "profile" than stations below that latitude.

**A.2. Use of hourly values instead of hourly means in older data**

Originally (*i.e.* more than ~150 years ago), magnetic measurements were eye-readings taken at discrete times. Magnetic data yearbooks (often containing meteorological data as well) giving data for each hour (usually on the local hour mark) were published as a reasonably compact representation of the variation of the various elements. After continuous recording was introduced by *Brooke* [1847], the sheer mass of data soon overwhelmed the observers and the yearbooks still contained only hourly values. *Schmidt* [1905] pointed out that hourly *means* would use the records more fully than just the instantaneous hourly values and would also "eliminate the accidental character of chance disturbances". Starting with the 1905 yearbook, *Schmidt* published hourly means for Potsdam [POT] (modern replacement station is now Niemegk [NGK]) near Berlin and soon most observers followed his lead, although for some it took quite some time (Chambon-la-Forêt [CLF] changed from hourly values to hourly means only in 1972). Owing to the higher variability of instantaneous values as compared to the smoother mean values, *IHV* is considerably higher (up to 50% for some stations) when computed from instantaneous values rather than from mean values. Using modern one-minute values we can readily create a data set with near instantaneous values spaced one hour apart as well as calculate hourly means from 60 one-minute values. At our urging, *Mursula and Martini* [2006] came to the same conclusion. Figure 23 shows *IHV* for NGK calculated from the hourly values (denoted $IHV_{01}$) and from the hourly means (denoted $IHV_{60}$).

As the Figure shows, in a first approximation, we have to multiply $IHV_{01}$ by 0.7065 to reduce the values to $IHV_{60}$. The importance of this reduction was not clear in our preliminary study of *IHV* [*Svalgaard et al.*, 2004]. For times when geomagnetic activity is low, the difference between hourly values and hourly means becomes smaller and $IHV_{01}$ approaches $IHV_{60}$. Applying a constant, average conversion factor between $IHV_{01}$ and $IHV_{60}$ will thus tend to slightly underestimate the $IHV_{60}$ calculated from $IHV_{01}$. This has the undesirable side effect of introducing a slight, and spurious, solar cycle dependence for the ratio between $IHV_{60}$ and $IHV_{01}$. A better fit is a power-law applied to the daily values of *IHV*. Inline Table A.2.a gives the parameters *a* and *b* for power-laws y = $a\ x^b$ and times of changeover from hourly values to hourly means that we have determined for the stations used. Some of these times are dictated by the timing of data gaps rather than positive knowledge of when the changeover actually took place. If no one-minute data were available, the approximate method used for Figure 23 is used.

[Table A.2.a]

| Obs. | *a* | *b* | Before | Note |
|---|---|---|---|---|
| POT | 1.1715 | 0.8668 | 1905 | Assumed the same as for NGK |
| WLH | 1.1715 | 0.8668 | 1912 | Assumed the same as for NGK |
| CLH | 1.1859 | 0.8756 | 1915 | Assumed the same as for FRD |
| VQS | 1.0415 | 0.9286 | 1915 | Assumed the same as for SJG |
| TUC | 1.1008 | 0.9049 | 1915 | |
| HON | 1.2595 | 0.8701 | 1915 | |
| API | 1.7190 | 0.6856 | 1929 | Noisy |
| KAK | 1.0890 | 0.9316 | 1955 | |
| TOK | 1.0890 | 0.9316 | 1913 | Assumed the same as for KAK |
| DBN | 0.7000 | 1.0000 | 1938 | Linear fit to SED |
| WIT | 0.7890 | 1.0000 | 1984 | Linear fit to NGK |
| ESK | 0.7119 | 1.0000 | 1919 | Multiply by 1.484 before 1932 (see text) |
| VLJ | 1.0312 | 0.9096 | 1938 | Assumed the same as for CLF |
| CLF | 1.0312 | 0.9096 | 1972 | |
| VSS | 0.8666 | 1.0000 | 1926 | Multiply by 1.45 in 1921/8-1925/3 (see text) |
| PIL | 0.8450 | 1.0000 | 1949 | Comparison with SJG |
| SVD | 0.7235 | 1.0000 | 1932 | Linear fit to ABG |

### A.3. Dealing with Extreme Activity

Indices like *am* and *aa*, that are derived from *K*-indices are capped at the top amplitude associated with $K = 9$. At times of great disturbance, *IHV* can be very large, exceeding even this maximum amplitude. An example of this is evident even in the monthly means in Figure 2. We wish to limit (or *cap*) *IHV* at a suitable maximum amplitude to avoid such extremes. Experiments show that a cap of 7 times the average *IHV* largely eliminates such chance outliers. If the difference is larger, it is set equal to the cap-value. This happens about 0.25% of the time. Figure 24 shows rotation averages of *Am* (black curve) compared to *IHV* from NGK (blue curve) [scaled to *Am* using eq. (4)] derived using the cap. The red curve (it is there, but almost always hidden behind the blue curve) shows what *IHV* would have been without the cap. It is clear that the cap is needed to make the blue curve match the black curve and to prevent the red "spikes".

### A.4. Station specific artifacts in archival data

### A.4.1. The Curious Case of Eskdalemuir

The 100 years of records from ESK could be an important source of *IHV* in the European-African longitude sector. Both *Mursula et al.* [2004] and *Clilverd et al.* [2005] analyzed *IHV* derived from ESK and found very small values in the early one third of the 20$^{th}$ century supporting their notion of a significant centennial change of geomagnetic activity since then. Examination of the original yearbooks from ESK (one of us [LS] visited the observatory on the occasion of its centenary) revealed that the data available from the WDCs has been processed to simulate hourly means centered on the half-hour. Up through 1917, ESK reported instantaneous values on the whole hour; thereafter, until 1932, ESK reported hourly means, but still centered on the whole hour. Figure 25 shows data from the WDC plotted together with data from the original yearbook for $30^{th}$ January 1924. It is unmistakable that the WDC data is simply interpolated between the whole-hourly data given in the yearbook. Such smoothing greatly diminishes the variability of the data to the point where *IHV* becomes ~35% too small. Creating a synthetic interpolated dataset from modern data shows that to 'undo' the effect of the smoothing, it is necessary (and it suffices) to multiply the *IHV* calculated from the smoothed data by 1.485. This removes the ESK anomaly and brings ESK into agreement with other stations [*Martini and Mursula*, 2006], but, of course, also invalidates conclusions based on the uncorrected ESK data (*e.g. Mursula et al.* [2004] and *Clilverd et al.* [2005]).

**A.4.2. Quality Control of WDC Data**

A large database invariably contains errors of many types: timing, calibration, sign, transcription, omission, and misunderstanding. Information *about* the data (metadata) is sorely lacking, especially for older data in the classical WDC data format. Fortunately, the data themselves can often be analyzed to bring to light many errors and allow correction of much of the data. Unfortunately, it is difficult to propagate such corrections back into the publicly available databases. This section details our experience with a typical case, Vassouras (VSS) near Rio de Janeiro. The observatory has been in continuous operation since 1915 and is important as the longest running station in its

739  longitude sector in the Southern Hemisphere. Figure 26 shows the diurnal variation of the

740  horizontal component through the years. It is evident that about half of the data is not in

741  the WDCs. The daily maximum occurs at $14.7^h$ UT or $11.8^h$ local time. The data from the

742  WDC is consistent with this, but only during 1957-1959 and 1998-present. At other times

743  the maximum occurs 3 hours earlier (after about 1925) or 3.5 hours earlier (before that).

744  The early WDC data thus seem to be hourly instantaneous values taken at the whole-hour

745  and then hourly means centered at the half-hour (some time after 1925) according to *local*

746  standard time (Brazil started to use standard time Jan. $1^{st}$ 1914) rather than UT. We

747  shifted the hourly data points to four hours later before 1926 and to three hours later for

748  1949-1997 except for the IGY-data 1957-1959 that has already been shifted in the WDC

749  data. It is unknown (to us) when the change from hourly values to hourly means took

750  place, although the time of the Second Polar Year 1932-33 would be a likely candidate.

751  The Tables in the yearbooks that give the hourly data values are conventionally *based* in

752  the sense that actual value of the field is *Field = Base ± Tabular Entry*. The sign of the

753  tabular entry is usually '+", but occasionally '-' is used, *e.g.* as evidenced in an indirect

754  way by this quote from a Batavia (BTV) yearbook "increasing numbers denote

755  decreasing easterly declination". Such subtlety is often lost during the data entry process

756  (partly because the sign used may change without warning for a given station at any

757  time). The base sometimes changes during a year as well. This also is often not caught at

758  data entry so that the data values entered are off by (usually) a multiple of 100 (*e.g.* H

759  component for the year 1957 and August 1959 for VSS, except the first three hours,

760  because of the local time to UT shift). Base and sign changes may be difficult to correct

761  because the data may have been reformatted later (maybe even at the WDC). This seems

762  to have happened to the VSS data, because the diurnal *maximum* at $11.8^h$ local time
763  occurs as a *minimum* during the interval January 1915 through May 1918, yet the base
764  value is the same on either side of the $31^{st}$ May 1918. We changed the sign of the tabular
765  entries and adjusted the base values before June 1918, removed spurious data, *e.g.* for the
766  $31^{st}$ November 1972 (in data from WDC Kyoto), corrected base offsets when they were
767  clearly wrong, and decided to completely omit data for April 1991 because the tabular
768  entries were in units of 10 nT rather than 1 nT (inferred from a ten-fold diminution of the
769  variability during that month), and still there were errors left as detailed in the discussion
770  of VSS in section B.6. The main point here is to emphasize that the data in the WDCs
771  contain errors that realistically can only be reliably corrected at or with cooperation from
772  the observing stations themselves as the necessary metadata may only be available locally
773  at the observatories or their managing institutions.

774  A few other examples further illustrate problems of data quality. The data for VQS is
775  given in local time rather than UT, and the data for the series POT-SED-NGK has the $7^{th}$
776  through the $10^{th}$ day of every January for many early years designated as "missing",
777  although the data is present in the yearbooks and in older versions of the electronic data.
778  Apparently, some recent "cleanup" was attempted with unintended consequences. For
779  years from 1900 through 1907, the WDCs at times have "Y2K" problems where there is
780  confusion about 2000-2007 and 1900-1907, as the century has no unique designation
781  within the "WDC Exchange Format". Attempts by the WDCs to rectify these various
782  problems have often led to introduction of other problems or to loss of data, and there is
783  no standard procedure for feedback from researchers to the WDCs.

## A.5. Secular Changes of the Earth's Main field

### A.5.1. Influence of Changing Corrected Geomagnetic Latitude

As Figure 6 shows, *IHV* for stations close to or polewards of 55º is very sensitive to changes in the stations' corrected geomagnetic latitude. *Mursula et al.* [2004] analyzed *IHV* derived from (among others) SOD (~63º) and *Clilverd et al.* [2005] analyzed *IHV* derived from (among others) LER (~59º). Both these stations are polewards of 55º and shouldn't be directly compared with results from lower latitudes. Figure 27 shows why. From 1900 to 2005, the corrected geomagnetic latitude of LER decreased from 59.35º to 57.98º, while SOD increased from 62.35º to 64.10º. Using eq.(3) we calculate *IHV* for these stations for every five years and plot the percentage change relative to their mean values over the time interval. The total change is +27.5% for SOD and -36.5% for LER. Owing to their lower latitude, the change for ESK is smaller (-6.5%) and for NGK is negligible (+0.5%) [part of the reason that NGK was chosen as reference station].

It is instructive to calculate *IHV* from the actual data for LER and SOD (the latter scaled by 0.2579 to match the mean of LER). Corrected geomagnetic latitude is determined using the International Geomagnetic Reference Field IGRF/DGRF models supported by the http://modelweb.gsfc.nasa.gov/models/cgm/cgm.html website. The lower panel in Figure 27 shows the ratio LER/(scaled SOD) for each Bartels rotation since 1926. The red line shows the ratio expected, using eq.(3), due to the changing latitudes: a steady decrease totaling 46%. We note good agreement and also that the effect is large. These stations should not be used without correcting for the secular change of latitude, or better, not used at all as representative for any purported global change on account of their latitude being too high. This invalidates the conclusions of *Mursula et al.* [2004] and *Clilverd et al.* [2005] regarding long-term changes of the solar wind.. For all stations ultimately used (as indicated in Table 1) we have verified that the changes of *IHV* due to secular changing corrected geomagnetic latitudes are small and of varying sign such as to be negligible in a composite time series.

**A.5.2. Influence of Decreasing Main Field**

The dipole moment of the Earth's main magnetic field has decreased significantly over the past centuries, influencing both the size of the magnetosphere and the conductivity of the ionosphere. *Glassmeier et al.* [2004] found that ring current perturbations (measured by the *IDV*-index) do not increase with decreasing dipole moment $M$, and suggest that the magnetic effect, $b$, of the polar electrojets (partly measured by the *IHV*-index) scales very weakly with $M$, *viz.* as $M^{1/6}$. If so, the 9% decrease of $M$ since 1850 would only result in a 1.5 % decrease of $b$ which would be too small to reliably detect. On the other hand, there is evidence [*Svalgaard and Cliver*, 2007] that the conductance of the midlatitude ionosphere has increased by 10% since 1840. This increase is reflected in the amplitude of the $S_R$ variation, but it is not clear what effect that has on nighttime geomagnetic activity. We therefore do not attempt to correct for the change in the main field. It is, however, an important and as yet unresolved question what the effect of a changing main field has on observed geomagnetic activity. In the present paper we assume that to first order the effect can be ignored. This assumption may not be valid, so all results are appropriately qualified.

**A.6. Missing Data During Storms**

828  There are some effects that can lead to an underestimation of *IHV*, such as data missing
829  because of the magnet moving too rapidly or the recording going off scale. A typical
830  example is the very brief magnetic storm of September 25$^{th}$, 1909, possibly the strongest
831  storm during the entire interval 1882-2006 [*Love*, 2006]. We suggest using other indices
832  like the *aa*-index to infer *IHV* for times with missing data as follows. For the interval of
833  27 Bartels rotations (~2 years) centered on the rotation containing the storm sudden
834  commencement, a 'sectorial' *aa*-index can be computed for each rotation as the average
835  for that rotation of the *aa*-index over the UT-intervals matching as closely as possible the
836  six-hour intervals used for the calculation of *IHV* for the six longitude sectors. Omitting
837  the data for the rotation with the storm (or missing data in general), the *aa* values are then
838  scaled (see Appendix B) to the average of *IHV* for each sector. Applying the scale factors
839  so gained, *aa* can now be scaled up to estimate *IHV* for the corresponding sectors (both
840  North and South) during the rotation with the storm-related missing data. Because each
841  storm is handled separately, correct calibration and long-term stability of the *aa* or other
842  indices are not essential. The estimate is crude, but is better than having no estimate at
843  all. For times where no other indices are available one could use a 'climatological' storm
844  profile. We have not yet carried out this procedure for the present investigation.

845  **A.7. Lack of Consistent Summer/Winter Difference**

846  Even though we have removed the dominant equinoctial contribution of the semiannual
847  variation we expect a residual semiannual variation due to axial and Russell-McPherron
848  effects, as well as a possible difference related to the seasons (Summer/Winter). Figure
849  28 shows the variation of *IHV* with month of year. The upper panel shows the annual
850  variation of *IHV* for the Northern Hemisphere (blue), Southern Hemisphere (red), and composite Equatorial (green) series for years 1940 to the present. The average of these three series is shown with a thick black curve. Below this curve we show (purple) the average annual variation of the full *IHV* series for years before 1940 where the data is sparser, especially for the Southern Hemisphere. In spite of the larger noise level, the same general variation is found in this subset as well, namely a superposition of an annual wave with extrema near aphelion and perihelion and a semiannual variation with extrema near the usual times.

The dotted curve shows the variation of the "raw" *IHV* (*i*.e. not corrected for the dipole tilt modulation). To better show the annual variation we have repeated the curves for yet another year in the right-hand portion of the Figure. The residual variation (after correction) is about 25%, consistent with several studies (*e.g. Svalgaard et al.* [2002] and pertinent references therein). Drawing on a result of section 6 we expect the residuals to be largely driven by $BV^2$. This seems to be the case as shown by the lower panel of Figure 28 that depicts the average annual variation of IMF *B* (blue), solar wind speed *V* (red), and the product $BV^2$ (thin black) relative to their mean values for 1965-2006. The heavy black curve shows a three-point running mean of the normalized $BV^2$. Well aware of the danger of over-interpreting noisy data, we suggest that the dominant variation of *B* is an annual wave with minimum near aphelion and maximum near perihelion. The amplitude of this annual wave is consistent with what we would expect from the variation of the distance from the Sun due to the eccentricity of the Earth's orbit. With this interpretation, most of the annual variation of *IHV* is explained and there seems to be little room for an intrinsic systematic local summer/winter difference.

## Appendix B.

### B.1. European-African Sector (15º E)

We start in the North. NGK replaced SED in 1932, so the first task is to bridge SED and NGK. The four stations VLJ, ABN, RSV, and DBN overlap the transition from SED to NGK. Inline Table B.1a gives the scale factors to apply to *IHV* derived from the station in the first column to SED and NGK, respectively. The ratio between these factors is the scale factor for normalizing SED to NGK; we adopt for this value the average (0.9349±0.0082) of the four stations.

[Table B.1a]

| From | SED | $R^2$ | Time | NGK | $R^2$ | Time | NGK/SED |
|---|---|---|---|---|---|---|---|
| VLJ | 1.2045 | 0.87 | 1923-1931 | 1.1399 | 0.77 | 1932-1936 | 0.9463 |
| ABN | 1.1196 | 0.94 | 1926-1931 | 1.0370 | 0.88 | 1932-1956 | 0.9263 |
| RSV | 1.0077 | 0.96 | 1927-1931 | 0.9233 | 0.89 | 1932-1978 | 0.9163 |
| DBN | 1.0103 | 0.85 | 1908-1931 | 0.9604 | 0.80 | 1932-1938 | 0.9506 |
| | | | | | | | 0.9349 |

The next step is to multiply SED by its scaling factor to NGK and join the scaled SED to the NGK series for a combined SED-NGK series. Then we regress the four stations against this combined series and arrive at the final set of scale factors shown in Table B.1b.

[Table B.1b]

| from | NGK | $R^2$ | Time |
|---|---|---|---|
| VLJ | 1.1399 | 0.85 | 1923-1936 |
| ABN | 1.0364 | 0.89 | 1926-1956 |
| RSV | 0.9233 | 0.90 | 1927-1978 |
| DBN | 0.9604 | 0.84 | 1908-1938 |
| SED | 0.9349 | Adopt | 1908-1931 |

Although RSV (Rude Skov) continued operation through 1981, the data after 1978 is too noisy to be of any use. Extension of a nearby electric trainline necessitated relocating the observatory to rural Brorfelde. RSV goes back to 1908 and VLJ to 1901. Their predecessor stations, COP and PSM operated from 1891 and 1883, respectively, and data exist, but not yet in electronic form.

There are three further dates of concern: the interruption of observations at the end of World War II, the start of the IGY (when some observatories improved instruments and/or reduction procedure), and the introduction of digital recording in the 1980s. To verify the stability of NGK from 1932 to the present, we compare with FUR, WNG, WIT, BFE, HAD, and CLF. Table B.1c gives the scale factors to NGK for these stations.

[Table B.1c]

| from | NGK | $R^2$ | Time |
|---|---|---|---|
| FUR | 1.1251 | 0.96 | 1940-2004 |
| WNG | 0.9269 | 0.97 | 1943-2004 |
| WIT | 1.0037 | 0.97 | 1939-1984 |
| BFE | 0.9403 | 0.95 | 1981-2004 |
| CLF | 1.1466 | 0.87 | 1936-2004 |
| HAD | 1.0823 | 0.89 | 1957-2004 |

There are no indications of any problems or systematic differences and the correlations are uniformly high, so we conclude that the calibration is stable and that all these stations support each other. Note that we did not include ESK in this group, because of its proximity to the auroral zone and the problems with its WDC data set described in section A.4.1.

There remains to join POT to the reference series. DBN data from 1903 through 1938 form the bridge between POT and NGK. The scale factors are given in Table B.1d from which we derive the scale factor from POT to NGK equal to (DBN→NGK/DBN→POT) = 0.9604/0.9819 = 0.9871.

[Table B.1d]

| from | POT | $R^2$ | Time | NGK | $R^2$ | Time |
|------|--------|------|-----------|--------|-------|-----------|
| DBN  | 0.9819 | 0.90 | 1903-1907 | 0.9604 | 0.80  | 1932-1938 |
| POT  |        |      |           | **0.9871** | *Adopt* | 1890-1907 |
| WLH  | 0.9451 | 0.81 | 1883-1895 | **0.9329** | *Adopt* | 1883-1895 |

Hourly values from WLH [kindly supplied by *J. Linthe*, personal communication, 2005] overlap with POT for 1890-1895. From the overlap between WLH and POT we compute the scale factor from WLH to NGK as (WLH→POT*POT→NGK) = 0.9451*0.9871 = 0.9329.

It would be highly desirable to verify and solidify these calibrations using the several other European stations observing at the time, but to date no other data exist in electronic form. We can now construct a composite series for the Northern Hemisphere European

Sector (IHV15N). The average standard deviation around the mean value is 1.4 nT (*IHV* has units of nT) or 3.8%.

Now the South. Suitable stations are HBK and CTO and its replacement station HER. Their scale factors to NGK are given in Table B.1e. Whether or not there is a real intrinsic difference in geomagnetic activity between the two hemispheres is not known and we shall leave it at that, normalizing to NGK.

[Table B.1e]

| from | NGK | $R^2$ | Time |
|---|---|---|---|
| CTO | 1.4342 | 0.76 | 1932-1940 |
| HER | 1.3505 | 0.80 | 1941-2004 |
| HBK | 1.2215 | 0.73 | 1973-2004 |

Finally, IHV15S is scaled to IHV15N. The scale factor is given in Table 5.

**B.2. Siberian-Indian Sector (75º E)**

We start in the North. We select ABG (digital data for 1924-2004) as reference station for this sector, then compute the average of all stations in the sector (becomes IHV75N), and finally scale that average to IHV15N. As with the question of intrinsic North/South difference we also force all sectors to (ultimately) the same level as NGK. We have already compensated for the (real) UT-difference between sectors (see section 3.2). Table B.2a gives the scale factors to ABG.

[Table B.2a]

| from | ABG | $R^2$ | Time |
|---|---|---|---|
|  |  |  |  |
| SVD | 1.1251 | 0.78 | 1930-1980 |
| TFS | 0.8190 | 0.81 | 1957-2001 |
| TKT | 0.9847 | 0.90 | 1957-1991 |
| NVS | 0.8592 | 0.82 | 1967-2003 |
| AAA | 0.9340 | 0.86 | 1963-2002 |
| ARS | 0.9009 | 0.79 | 1973-2002 |

Finally, IHV75N is scaled to IHV15N The scale factor for IHV75N to IHV15N is given in Table 5.

Now the South. Suitable stations are AMS, CZT, and PAF, although PAF is at an uncomfortably high corrected geomagnetic latitude (-58º) and its scaling to AMS is therefore decidedly non-linear ($IHV_{AMS} = 2.114\ IHV_{PAF}^{0.634}$; $R^2 = 0.85$). We then construct IHV75S as the mean of AMS and the scaled CZT and PAF. Table B.2b gives the scale factor for CZT to AMS.

[Table B.2b]

| from | AMS | $R^2$ | Time |
|---|---|---|---|
| CZT | 0.8420 | 0.83 | 1974-2004 |

Finally IHV75S is scaled to IHV15N. The scale factor is given in Table 5.

**B.3. Japanese-Australian Sector (130º E)**

We start in the North. We select KAK as reference station for this sector, deduce the scaling factors for the other stations in this sector, then compute the average of the scaled

952  values (becomes IHV130N), and finally scale that average to IHV15N. Suitable stations

953  are MMB, KNY, and SSH. Table B.3a gives the scale factors to KAK for these stations.

954  [Table B.3a]

| From | KAK | $R^2$ | Time |
|------|--------|-------|-----------|
| MMB  | 0.8448 | 0.98  | 1957-2006 |
| KNY  | 0.9526 | 0.96  | 1958-2006 |
| SSH  | 0.8148 | 0.92  | 1933-2002 |
| TOK  | 0.6950 | 0.50  | 1897-1912 |

955

956  Data for Tokyo (TOK) was kindly supplied by *Takashi Koide* [personal communication,

957  2005]. TOK was "bridged" to KAK using IHV15N. Undigitized data for this sector exist

958  to eventually improve this bridge. The scale factor for the composite IHV130N to

959  IHV15N is given in Table 5.

960  Now the South. The Australian stations all fall in this sector. We select GNA as reference

961  station (1957-2004). The earliest available data is from WAT (1919-1958). TOO (1924-

962  1979) can then serve as bridge between WAT and GNA. For modern data, CNB (1981-

963  2004) serves as a check on GNA. Scale factors are given in Table B.3b. The scale factor

964  from WAT to GNA is (TOO→GNA/TOO→WAT) = 0.8041/0.8549 = 0.9406.

965  [Table B.3b]

| from | WAT    | $R^2$ | Time      | GNA    | $R^2$ | Time      | GNA/WAT |
|------|--------|-------|-----------|--------|-------|-----------|---------|
| TOO  | 0.8549 | 0.84  | 1924-1958 | 0.8041 | 0.83  | 1957-1979 | 0.9406  |
| CNB  |        |       |           | 0.8468 | 0.79  | 1981-2004 |         |
| WAT  |        |       |           | 0.9406 | *adopt* |         |         |

966

Finally IHV130S is scaled to IHV15N. The scale factor is given in Table 5.

**B.4. Mid-Pacific Sector (200º E)**

We start in the North. We select HON as reference station for this sector. There are data for a few years from MID. We include that data to compare with HON. As Table B.4a show, the agreement between MID and HON is good. Finally IHV200N is scaled to IHV15N. The scale factor is given in Table 5.

[Table B.4a]

| From | HON | $R^2$ | Time |
|------|--------|------|-----------|
| MID  | 1.0063 | 0.92 | 2000-2002 |

Now the South. We select API (1922-2004) as reference station and scale AML (1957-1978), EYR (1978-2004), and PPT (1968-2004) to API. The scale factors are given in Table B.4b. Finally, IHV200S is scaled to IHV15N. The scale factor is given in Table 5.

[Table B.4b]

| From | API | $R^2$ | Time |
|------|--------|------|-----------|
| AML  | 0.7839 | 0.79 | 1957-1978 |
| EYR  | 0.7758 | 0.77 | 1978-2004 |
| PPT  | 1.0455 | 0.85 | 1968-2004 |

**B.5. Pacific West Coast Sector (240º E)**

We start in the North. We select TUC (1910-2004) as reference station for this sector because of its very long series of observations. BOU (1967-2004), FRN (1983-2004), and

983 VIC (1957-2004) comprise the remaining stations of the sector. TEO would have been
984 ideal, but is very noisy ($R^2$ for correlation with TUC is only 0.24). Scale factors are given
985 in Table B.5a. A power law for VIC is a slightly better fit: TUC = 1.6102 VIC $^{0.849}$ ($R^2$ =
986 0.85) and is used instead of the linear fit.

987 [Table B.5a]

| From | TUC | $R^2$ | Time |
|---|---|---|---|
| VIC | 0.9234 | 0.82 | 1957-2004 |
| BOU | 1.0345 | 0.94 | 1967-2004 |
| FRN | 1.0626 | 0.89 | 1983-2004 |

988

989 Finally IHV240N is scaled to IHV15N. The scale factor is given in Table 5.

990 Now the South. There are really no stations with available data. Easter Island (EIC)
991 would be ideal, but only 15 days of data (in 1964) have been found, although yearbooks
992 or data may be available for other years. IAGA Resolution 8 (1979) urged establishment
993 of an observatory on Easter Island, and the French IPGP is planning such a station under
994 the INTERMAGNET program. In anticipation hereof and for completeness we include
995 EIC, becoming IHV240S. IHV240S is finally scaled to IHV15N. The scale factor is
996 given in Table 5.

997 **B.6. The Americas Sector (290º E)**

998 We start in the North. We select FRD (1956-2004) as reference station for this sector. We
999 need SJG (1926-2004) to serve as a strong bridge between CLH (1901-1956) and FRD
1000 (1956-2004). CLH in turn bridges the gap between VQS (1903-1924) and SJG. The data in the WDC for VQS was given in local standard time rather than UT and had to be shifted appropriately. Scale factors are given in Table B.6a. The scale from CLH to FRD is (CLH→SJG/FRD→SJG) = 0.7323/0.7869 = 0.9306. In a similar manner we derive the scale factors for all the stations to FRD shown in the last column of the table.

[Table B.6a]

| from | SJG | $R^2$ | Time | CLH | $R^2$ | Time | FRD |
|---|---|---|---|---|---|---|---|
| CLH | 0.7323 | 0.85 | 1926-1956 | | | | 0.9306 |
| FRD | 0.7869 | 0.85 | 1956-2004 | | | | 1.0000 |
| VQS | | | | 1.4150 | 0.84 | 1903-1924 | 1.3168 |
| SJG | | | | 1.3655 | 0.85 | 1926-1956 | 1.2708 |

Finally IHV290N is scaled to IHV15N. The scale factor is given in Table 5.

Now the South. We select VSS (1915-2006) as reference station. Data for 1921 August - 1926 are too low compared to IHV290N and have been scaled up by a factor of 1.45. No metadata is available yet to help explain the reason for this discrepancy. PIL (1940-1985) and TRW (1957-2004) supply additional data, filling gaps in the series for VSS. Scale factors are given in Table B.6b.

[Table B.6b]

| From | VSS | $R^2$ | Time |
|---|---|---|---|
| PIL | 1.0238 | 0.58 | 1949-1985 |
| TRW | 0.9863 | 0.72 | 1957-2004 |

Because both VSS and TRW have many data gaps, but one often has data while the other one does not, we construct a combined VSS-TRW dataset (scaled to VSS) and use AIA (1957-2004), LQA (1964-1981), SGE (1975-1982), ARC (1978-1995), PST (1994-2004), and LIV (1996-2005) to fill in the holes. Scale factors to the combined VSS-TRW dataset are given in Table B.6c.

[Table B.6c]

| From | VSS-TRW | $R^2$ | Time |
|------|---------|-------|------|
| AIA  | 0.8585  | 0.74  | 1957-2004 |
| LQA  | 0.9904  | 0.75  | 1964-1981 |
| SGE  | 0.7871  | 0.46  | 1975-1982 |
| ARC  | 0.9921  | 0.75  | 1978-1995 |
| PST  | 1.0006  | 0.92  | 1994-2004 |
| LIV  | 0.8932  | 0.90  | 1996-2005 |

Finally IHV290S is scaled to IHV15N. The scale factor is given in Table 5.

**B.7. Equatorial Stations**

A selection of stations close to (within 9º of latitude) the geomagnetic equator was evaluated for suitability. Scaling factors to NGK are given in Table B.7a. The decrease in correlation is due to differences in longitude and thus in local time as we progress around the globe (see section 4.2). The lowest panel of Figure 7 compares the 13-rotation running mean of a composite of the equatorial stations and the global composite discussed in section 4.2. We conclude that *IHV* can be reliably derived even this close to the equator. Figure S13 in the Electronic Supplement shows the detailed composite *IHV*
for the Equatorial stations.

[Table B.7a]

| From | NGK | $R^2$ | Time |
|------|--------|-------|-----------|
| BNG  | 1.1467 | 0.76  | 1955-2003 |
| AAE  | 1.0607 | 0.78  | 1958-2004 |
| TRD  | 1.3365 | 0.65  | 1957-1999 |
| ANN  | 1.2659 | 0.65  | 1964-1999 |
| GUA  | 1.4146 | 0.46  | 1957-2004 |
| HUA  | 1.0917 | 0.48  | 1955-2004 |

The equatorial stations were not included in deriving the global composite *IHV*, because we want to compare *IHV* to the mid-latitude range indices. In the future one might contemplate including the equatorial *IHV*.

**Acknowledgments**

Geomagnetic data has been downloaded from the World Data Centers for Geomagnetism in Kyoto, Japan, and Copenhagen, Denmark. The research results presented in this paper rely on the data collected at magnetic observatories worldwide, and we thank the national institutions that support them. We also recognize the role of the INTERMAGNET program in promoting high standards of magnetic observatory practice. We thank the many people who have helped us with collection of data and metadata, especially J. Linthe, J. Love, T. Koide, H. Nevanlinna, Z. Kobylinski, J. Matzka, and H. Coffey. We also thank an anonymous reviewer for extensive and constructive comments.

van Sabben, D. (ed.), (1977), Geomagnetic data 1977: Indices, Rapid Variations, Special Intervals, *IAGA Bulletin 32h.*, IUGG Publications, Paris.

Wilcox, J. M. and P. H. Scherrer (1972), Annual and solar magnetic cycle variations in the interplanetary magnetic field 1926-1971, *J. Geophys. Res.*, *77*, 5385.

───────────────

L. Svalgaard, Easy Toolkit, Inc, 6927 Lawler Ridge, Houston, TX 77055, USA. (leif@leif.org)

E. W. Cliver, Air Force Research Laboratory, Hanscom AFB, MA 01731, USA. (Edward.Cliver@hanscom.af.mil)

**Table Captions**

**Table 1.** Geomagnetic observatories used in the present study. Listed are geographic longitude and latitude, UT time of local geographic midnight, the number of hours to skip to reach the six-hour interval used to calculate *IHV* (see section 3.1), corrected geomagnetic latitude (for the middle of the operating interval), the operating years interval, and the interval for which digital data were available at the time of writing.

**Table 2.** Yearly values of composite *IHV*, *B* derived from the *IDV*-index, *V* calculated using eq.(11), and *B* and *V* observed by spacecraft. *B* in nT and *V* in km/s. Values for 2006 and 2007 are preliminary only, based on incomplete data.

**Table 3.** Geomagnetic observatories with long series of data that may be useful for constructing *IHV*-indices. If a station stopped observing, the next column(s) may give the replacement station(s) (if any). For many stations there are data even earlier than given here, *e.g.* Paris and Munich. The coordinates given in the first column are geographic longitude and latitude.

**Figure Captions**

**Figure 1.** Variation of the geomagnetic elements at Fredericksburg May 11-15, 1999 (UT). The "effective" noon is marked with a green line on May 15. The red boxes

indicate the six hours around midnight where the regular variation is absent or minimal. These intervals are used to define the *IHV*-index. May 11 is a good example of a day with very little activity. It is, in fact, the famous day where "the solar wind disappeared" [*e.g. Jordanova et al.*, 2001]. The solar wind momentum flux was only 1% of its usual value and the magnetosphere diameter was five times larger than normal. The interplanetary magnetic field was not affected and had its usual properties. The variability of $S_R$ is clearly seen by comparing May 11 and May 15.

**Figure 2.** Comparison of monthly means of the "raw" *IHV*-index (blue) calculated for the H-component at Fredericksburg and the *Am2*-index (red) for the interval 1970-76. The year-labels on the abscissa mark the beginning of each year. The thin pink curve shows *IHV* scaled down by 0.7475 for direct comparison with *Am2*. Note that *Am2* is the average *am*-index for two three-hour intervals.

**Figure 3.** Correlation between yearly averages of *IHV* calculated for KAK for the interval 1965-2006 and the quantity $BV^2$ (see section 6.2) as a function of the number of hours from $0^h$ UT to skip before calculating *IHV*. Blue curve is for the H-component, green for the Z-component, and pink for the D-component. The triangle shows the correlation for the number of "skip hours" adopted for this station (12 in this case).

**Figure 4.** Variation of the *S*-function (bottom panel) and of "raw" *IHV* (top panel) with month of year and Universal Time calculated for all the stations in Table 1 for *all* data available for each station. The *IHV* values for a given station were assigned to the Universal Time of local midnight. All values were divided by the average values for each station. The color coding over the ~40% variation is chosen such that purple to red represents low to high values.

**Figure 5.** Variation of *IHV* with corrected geomagnetic latitude. Average *IHV* over the interval 1996-2003 for each station with data in that interval are plotted. A few "outliers" (SIL, KRC, QSB, GLM, and KSH) are shown with small circles. Local induction effects may be responsible for these stations having about 25% higher *IHV*. The red curve shows a model fit to the larger circles as described in the text.

**Figure 6.** Bartels rotation means of *IHV* for BFE versus NGK for 1982-2004. The scale factor is derived as the slope of the regression line constrained to go through the origin. A single outlier marked with a large circle is not included in the fit.

**Figure 7.** (Upper) Plot of a portion (years 1990-2001) of all the individual data series that went into the composite series. Northern sectors are shown in black while southern sectors are shown in red. (Middle) Plot of the full series for years 1883-2006 (grey curve) overlain by its 13-rotation running mean (heavy black curve). The curve before 1890 is based on preliminary data from BTV and WLH. (Lower) Plot of 13-rotation running means of the composite *IHV* (blue) and *IHV* derived from Equatorial stations (red).

**Figure 8.** Relationship between Bartels rotation means of *Am* (freed for dipole tilt effect) and composite *IHV* for the interval 1959-2003.

1253

1254 **Figure 9.** (Upper panels) Bartels rotation averages of proxy values of *Am* calculated
1255 using eq.(4) (blue curve) and observed (red curve). Both datasets have been freed from
1256 the effect of the dipole tilt (section 3.2). The bottom panel shows the entire datasets
1257 overlain by their 13-rotation running means.

1258

1259 **Figure 10.** Relationship between Bartels rotation means of *Ap* (freed for dipole tilt effect)
1260 and composite *IHV* for the interval 1932-2004.

1261

1262 **Figure 11.** (Upper) Sample Bartels rotation averages of proxy values of *Ap* calculated
1263 using eq.(5) (blue curve) and observed (red curve). Both datasets have been freed from
1264 the effect of the dipole tilt (section 3.2). (Lower) Shows the entire datasets overlain by
1265 their 13-rotation running means.

1266

1267 **Figure 12.** Relationship between Bartels rotation means of *Aa* (freed for dipole tilt effect)
1268 and composite *IHV* for the interval 1980-2004.

1269

1270 **Figure 13.** (Upper) Sample Bartels rotation averages of proxy values of *Aa* calculated
1271 using eq.(6) (blue curve) and observed (red curve). Both datasets have been freed from
1272 the effect of the dipole tilt (section 3.2). (Lower) Shows the entire datasets overlain by
1273 their 13-rotation running means.

1274

**Figure 14.** Difference between observed and calculated values of Bartels rotation means of *Aa* showing the upward jump in 1957.

**Figure 15.** Relationship between Bartels rotation means of $BV_o^2$ and composite *IHV* for the interval 1965-2005.

**Figure 16.** Sample Bartels rotation averages of proxy values of $BV_o^2$ calculated using eq.(7) (blue curve) and observed (red curve).

**Figure 17.** 13-rotation running means of $BV_o^2$, calculated (blue curve) and observed (red curve). Areas of consistent disagreement are marked by ovals. These occur every other time when the Rosenberg-Coleman effect is large (amplitude on arbitrary scale given by green curve).

**Figure 18.** Relationship between yearly means of $BV_o^2$ and composite *IHV* for the interval 1965-2005. Years affected by the 22-year cycle are shown as open circles and are not included in the fit.

**Figure 19.** Yearly values of *B* (nT) derived from the *IDV*-index (eq.(10), upper blue curve) and *V* calculated using eq.(11) (lower blue curve). *V* is plotted as $V_o = V/100$ km/s. *B* and *V* observed in Space are shown in red.

**Figure 20.** Yearly values of $BV_o$ (blue curve) calculated from $B$ derived from the *IDV*-index (eq.(10)) and $V$ derived from the IHV-index (eq.(11)) compared to $BV_o$ (green curve) calculated from the range of diurnal variation of the horizontal component in the polar caps. $BV_o$ calculated from $B$ and $V$ observed in Space is shown in red.

**Figure 21.** The diurnal variation of the Y-component of the geomagnetic field at SBA (red, crosses), HUT (blue crosses; for 1902.5-1903.5), GJO (blue circles; for 1904), and GDH (red circles) all in a local coordinate system where the X-axis coincides with the average direction of the H-component. The curves have been shifted in time to have the same phase in each hemisphere, but out of phase between hemispheres. This is simply a presentation device to avoid having the curves crowd on top of each other. For SBA and GDH, modern data was used for years with approximately the same sunspot activity as during 1903-1904 as described in the text.

**Figure 22.** Diurnal variation of unsigned hourly differences (between one hour and the next) for the H-component as a function of local time shown for corrected geomagnetic latitude bands for all available stations during 1996-2003 (color-coded from red at the equator through green at midlatitudes to blue and black in the polar regions). A band contains all stations from both hemispheres described in section 3.3. The time extends over two days to position the six-hour midnight interval used for *IHV* in the middle of the Figure.

**Figure 23.** (Upper) Monthly means of *IHV* for Niemegk (NGK) 1996-2002. The heavy red curve shows *IHV* calculated from true hourly means (calculated as the mean of 60 one-minute values of the H-component). The blue curve shows *IHV* calculated from a single one-minute value taken each hour on the hour. The thin red curve shows the blue curve scaled down by the coefficient determined by the linear regression shown in the lower panel. (Lower) Correlation between the monthly means of *IHV* shown in the upper panel calculated from hourly means ($IHV_{60}$) versus calculated from hourly values (one-minute averages taken once an hour, $IHV_{01}$).

**Figure 24**. Rotation averages of *Am* (black curve) compared to *IHV* from NGK (blue curve) [scaled to *Am* using eq. (4)] derived using the cap. The red curve (almost always hidden behind the blue curve) shows what *IHV* would have been without the cap.

**Figure 25.** The variation of geomagnetic components X, Y, and Z on 30 January 1924 for ESK plotted using the hourly values supplied by the WDCs (blue diamonds) and given in the original observatory yearbook (red squares). It is unmistakable that the WDC data is simply interpolated between the whole hourly data given in the yearbook.

**Figure 26.** The diurnal variation of the horizontal component through the years for Vassouras (VSS) near Rio de Janeiro. The observatory has been in continuous operation since 1915 and is important as the longest running station in its longitude sector in the Southern Hemisphere. The plot is a contour-plot of the variation of the H-component about its daily mean as a function of the hour as given in the WDC-data (the "nominal"

hour). Colors from purple/blue to orange/red signify the range from low (negative) to high (positive) values. White areas show where data is missing from the WDC archive.

**Figure 27.** (Upper) Percentage change relative to the mean values over 1900-2005 of *IHV* expected for LER, SOD, ESK, and NGK [using eq.(3)] resulting from actual changes in corrected geomagnetic latitude for these stations.. (Lower) The ratio LER/(scaled SOD) for each Bartels rotation since 1926 of calculated *IHV* from the actual data for LER and SOD (the latter scaled by 0.2579 to match the mean of LER). The red line shows the ratio expected (from eq.(3)) due solely to the changing latitudes.

**Figure 28.** (Upper) Annual variation of *IHV* for the composite Northern Hemisphere (blue), Southern Hemisphere (red), and Equatorial (green) series for years 1940 to the present. The average of these three series is shown with a thick black curve. Below this curve we show (purple curve with open circles) the average annual variation of the full *IHV* series for years before 1940 where the data is sparser, especially for the Southern Hemisphere. The dotted curve shows the variation of the "raw" *IHV* (*i.e.* not corrected for the dipole tilt). To better show the annual variation we have repeated the curves for yet another year in the right-hand portion of the Figure. (Lower) Average annual variation of IMF *B* (blue), solar wind speed *V* (red), and the product $BV^2$ (thin black) relative to their mean values for 1965-2006. The heavy black curve shows a three-point running mean of normalized $BV^2$. It would seem that most of the annual variation of *IHV* can be explained simply as variation of the driving $BV^2$.

1366
1366| IAGA | Name | GG Long | GG Lat | Midnight | Skip | CGM Lat | From | To | From | To |
|------|------|---------|--------|----------|------|---------|------|------|------|------|
| VLJ | Val Joyeux | 2.0 | 48.8 | 23.9 | 21 | 44.9 | 1900 | 1936 | 1923 | 1936 |
| CLF | Chambon-la-Foret | 2.3 | 48.1 | 23.8 | 21 | 44.0 | 1935 | 2005 | 1936 | 2005 |
| DBN | De Bilt, Nederland | 5.2 | 52.1 | 23.7 | 20 | 48.5 | 1903 | 1938 | 1903 | 1938 |
| WIT | Wittingen | 6.8 | 52.8 | 23.5 | 20 | 49.2 | 1938 | 1984 | 1938 | 1984 |
| WLH | Wilhelmshafen | 8.2 | 53.5 | 23.5 | 20 | 50.8 | 1883 | 1911 | 1883 | 1895 |
| WNG | Wingst | 9.1 | 53.7 | 23.4 | 20 | 50.1 | 1939 | 2006 | 1943 | 2003 |
| FUR | Furstenfeldbruck | 11.3 | 48.2 | 23.2 | 20 | 43.5 | 1939 | 2006 | 1940 | 2004 |
| BFE | Brorfelde | 11.7 | 55.4 | 23.2 | 20 | 51.8 | 1981 | 2006 | 1981 | 2004 |
| RSV | Rude Skov | 12.5 | 55.5 | 23.2 | 20 | 51.9 | 1907 | 1981 | 1927 | 1981 |
| NGK | Niemegk | 12.7 | 52.1 | 23.2 | 20 | 48.0 | 1932 | 2006 | 1932 | 2004 |
| SED | Seddin | 13.0 | 52.3 | 23.1 | 20 | 48.2 | 1908 | 1931 | 1908 | 1931 |
| POT | Potsdam | 13.1 | 52.4 | 23.1 | 20 | 48.3 | 1890 | 1907 | 1890 | 1907 |
| TSU | Tsumeb | 17.6 | -19.2 | 22.8 | 20 | -29.4 | 1964 | 2006 | 1964 | 2004 |
| CTO | Cape Town | 18.5 | -34.0 | 22.8 | 20 | -41.5 | 1932 | 1940 | 1932 | 1940 |
| BNG | Bangui | 18.6 | 4.4 | 22.8 | 20 | -8.3 | 1952 | 2006 | 1955 | 2003 |
| HER | Hermanus | 19.2 | -34.4 | 22.7 | 20 | -41.8 | 1941 | 2006 | 1941 | 2004 |
| HBK | Hartebeesthoek | 27.7 | -25.9 | 22.2 | 19 | -27.1 | 1973 | 2006 | 1973 | 2004 |
| AAE | Addis Ababa | 38.8 | 9.0 | 21.4 | 18 | -0.2 | 1958 | 2006 | 1958 | 2004 |
| TFS | Tbilisi | 44.7 | 42.1 | 21.0 | 18 | 36.8 | 1938 | 2006 | 1957 | 2001 |
| CZT | Crozet | 51.9 | -46.4 | 20.5 | 18 | -53.2 | 1974 | 2004 | 1974 | 2004 |
| ARS | Arti | 58.6 | 56.4 | 20.1 | 17 | 51.7 | 1973 | 2006 | 1973 | 2002 |
| SVD | Sverdlovsk | 61.1 | 56.7 | 19.9 | 17 | 51.9 | 1929 | 1980 | 1930 | 1980 |
| TKT | Tashkent | 69.6 | 41.3 | 19.4 | 17 | 36.0 | 1883 | 1991 | 1957 | 1991 |
| PAF | Port aux Francais | 70.3 | -49.4 | 19.3 | 17 | -58.5 | 1957 | 2006 | 1957 | 2004 |
| ABG | Alibag | 72.9 | 18.6 | 19.1 | 16 | 11.3 | 1904 | 2006 | 1925 | 2004 |
| AAA | Alma-Ata | 76.9 | 43.3 | 18.9 | 16 | 37.9 | 1963 | 2002 | 1963 | 2002 |
| TRD | Trivandrum | 77.0 | 8.5 | 18.9 | 16 | 0.0 | 1854 | 2006 | 1957 | 1999 |
| AMS | Martin de Vivies | 77.6 | -37.8 | 18.8 | 16 | -46.5 | 1981 | 2006 | 1981 | 2004 |
| NVS | Novosibirsk | 82.9 | 55.0 | 18.5 | 15 | 49.9 | 1967 | 2006 | 1967 | 2005 |
| ANN | Annamalainagar | 79.7 | 11.4 | 18.7 | 16 | 3.0 | 1957 | 1999 | 1964 | 1999 |
| LRM | Learmonth | 114.1 | -22.2 | 16.4 | 13 | -33.4 | 1988 | 2006 | 1990 | 2004 |
| WAT | Watheroo | 115.9 | -30.3 | 16.3 | 13 | -42.7 | 1919 | 1959 | 1919 | 1958 |
| GNA | Gnangara | 116.0 | -31.8 | 16.3 | 13 | -44.4 | 1957 | 2006 | 1957 | 2002 |
| SSH | She-Shan | 121.2 | 31.1 | 15.9 | 13 | 24.0 | 1932 | 2006 | 1932 | 2002 |
| KNY | Kanoya | 130.9 | 31.4 | 15.3 | 13 | 24.1 | 1958 | 2006 | 1958 | 2006 |
| ASP | Alice Springs | 133.9 | -23.8 | 15.1 | 13 | -34.2 | 1992 | 2006 | 1992 | 2004 |
| TOK | Tokyo | 139.7 | 35.8 | 14.7 | 12 | 26.7 | 1897 | 1912 | 1897 | 1912 |
| KAK | Kakioka | 140.2 | 36.2 | 14.7 | 12 | 28.7 | 1913 | 2006 | 1913 | 2006 |
| MMB | Memambetsu | 144.2 | 43.9 | 14.4 | 12 | 36.5 | 1950 | 2006 | 1957 | 2006 |
| GUA | Guam | 144.9 | 13.6 | 14.3 | 11 | 5.4 | 1957 | 2006 | 1957 | 2004 |
| TOO | Toolangi | 145.5 | -37.5 | 14.3 | 11 | -48.6 | 1919 | 1979 | 1924 | 1979 |
| CNB | Canberra | 149.4 | -35.3 | 14.0 | 11 | -45.7 | 1979 | 2006 | 1979 | 2004 |
| EYR | Eyrewell | 172.4 | -43.4 | 12.5 | 10 | -50.2 | 1978 | 2006 | 1978 | 2004 |
| AML | Amberley | 172.7 | -43.2 | 12.5 | 10 | -49.9 | 1929 | 1977 | 1957 | 1977 |
| MID | Midway | 182.6 | 28.2 | 11.8 | 9 | 24.7 | 2000 | 2002 | 2000 | 2002 |
| API | Apia | 188.2 | -13.8 | 11.5 | 8 | -16.0 | 1905 | 2006 | 1922 | 2004 |
| HON | Honolulu | 201.9 | 21.3 | 10.5 | 8 | 21.7 | 1902 | 2006 | 1902 | 2004 |
| PPT | Pamatai | 210.4 | -17.6 | 10.0 | 7 | -16.3 | 1968 | 2006 | 1968 | 2004 |
| VIC | Victoria | 236.6 | 48.5 | 8.2 | 5 | 54.2 | 1956 | 2006 | 1957 | 2004 |
| FRN | Fresno | 240.3 | 37.1 | 8.0 | 5 | 43.6 | 1982 | 2006 | 1983 | 2004 |
| TUC | Tucson | 249.2 | 32.2 | 7.4 | 4 | 39.9 | 1909 | 2006 | 1909 | 2002 |
| BOU | Boulder | 254.8 | 40.1 | 7.0 | 4 | 48.5 | 1964 | 2006 | 1967 | 2004 |
| FRD | Fredericksburg | 282.6 | 38.2 | 5.2 | 1 | 50.3 | 1956 | 2006 | 1956 | 2004 |
| CLH | Cheltenham | 283.2 | 38.7 | 5.1 | 1 | 50.8 | 1901 | 1956 | 1901 | 1956 |
| HUA | Huancao | 284.7 | -12.0 | 5.0 | 2 | 1.2 | 1922 | 2006 | 1922 | 2004 |
| SJG | San Juan, PR | 293.8 | 18.1 | 4.4 | 1 | 30.0 | 1926 | 2006 | 1926 | 2004 |
| LQA | La Quiaca | 294.4 | -22.1 | 4.4 | 1 | -9.8 | 1920 | 1983 | 1968 | 1981 |
| VQS | Vieques | 294.5 | 18.3 | 4.4 | 1 | 30.1 | 1903 | 1924 | 1903 | 1924 |
| TRW | Trelew | 294.7 | -43.3 | 4.4 | 1 | -32.9 | 1957 | 2006 | 1957 | 2004 |
| AIA | Argentine Islands | 295.7 | -65.3 | 4.3 | 1 | -50.3 | 1957 | 2006 | 1957 | 2004 |
| PIL | Pilar | 296.1 | -31.7 | 4.3 | 1 | -17.6 | 1905 | 2006 | 1941 | 1985 |
| LIV | Livingstone Isl. | 299.6 | -62.7 | 4.0 | 1 | -48.0 | 1997 | 2006 | 1997 | 2005 |
| ARC | Arctowski | 301.5 | -62.2 | 3.9 | 1 | -47.6 | 1978 | 1995 | 1978 | 1995 |
| PST | Port Stanley | 302.1 | -51.7 | 3.9 | 0 | -38.1 | 1994 | 2006 | 1994 | 2004 |
| VSS | Vassouras | 316.3 | -22.4 | 2.9 | 23 | -14.7 | 1915 | 2006 | 1915 | 2004 |
| SGE | South Georgia | 324.0 | -54.5 | 2.4 | 22 | -44.4 | 1975 | 1982 | 1975 | 1982 |

| HAD | Hartland | 355.5 | 51.0 | 0.3 | 21 | 48.2 | 1957 | 2006 | 1957 | 2004 |
| ABN | Abinger | 359.6 | 51.2 | 0.0 | 21 | 48.0 | 1925 | 1958 | 1926 | 1956 |
| ESK | Eskdalemuir | 356.2 | 55.3 | 0.3 | 21 | 53.1 | 1908 | 2006 | 1911 | 2004 |
| LER | Lerwick | 358.8 | 60.1 | 0.1 | 21 | 58.1 | 1923 | 2006 | 1926 | 2004 |
| SOD | Sodankylä | 26.6 | 67.4 | 22.2 | 19 | 63.6 | 1914 | 2006 | 1914 | 2004 |

**Table 1.** Geomagnetic observatories used in the present study. Listed are geographic longitude and latitude, UT time of local geographic midnight, the number of hours to skip to reach the six-hour interval used to calculate *IHV* (see section 3.1), corrected geomagnetic latitude (for the middle of the operating interval), the operating years interval, and the interval for which digital data were available at the time of writing.

| | year | HV | DV | B obs | V calc | V obs | | year | HV | DV | | B obs | V obs |
|---|---|---|---|---|---|---|---|---|---|---|---|---|---|
| 1374 | year | HV | B DV | B obs | V calc | V obs | 1417 | 1932.5 | 35.15 | 5.67 | | 471 | |
| 1375 | 1890.5 | 23.01 | 5.47 | | 365 | | 1418 | 1933.5 | 31.48 | 5.53 | | 445 | |
| 1376 | 1891.5 | 31.13 | 6.15 | | 419 | | 1419 | 1934.5 | 27.13 | 5.53 | | 405 | |
| 1377 | 1892.5 | 39.15 | 7.69 | | 431 | | 1420 | 1935.5 | 30.42 | 5.87 | | 423 | |
| 1378 | 1893.5 | 32.35 | 6.90 | | 406 | | 1421 | 1936.5 | 30.45 | 6.29 | | 409 | |
| 1379 | 1894.5 | 38.47 | 7.92 | | 421 | | 1422 | 1937.5 | 34.84 | 7.43 | | 409 | |
| 1380 | 1895.5 | 33.99 | 6.59 | | 428 | | 1423 | 1938.5 | 39.21 | 8.08 | | 421 | |
| 1381 | 1896.5 | 34.56 | 6.62 | | 431 | | 1424 | 1939.5 | 42.07 | 7.61 | | 452 | |
| 1382 | 1897.5 | 26.74 | 6.37 | | 374 | | 1425 | 1940.5 | 40.31 | 7.39 | | 447 | |
| 1383 | 1898.5 | 30.53 | 5.93 | | 422 | | 1426 | 1941.5 | 42.41 | 7.45 | | 459 | |
| 1384 | 1899.5 | 26.97 | 5.54 | | 403 | | 1427 | 1942.5 | 39.82 | 6.46 | | 475 | |
| 1385 | 1900.5 | 19.63 | 5.02 | | 341 | | 1428 | 1943.5 | 43.67 | 6.32 | | 507 | |
| 1386 | 1901.5 | 16.65 | 4.66 | | 312 | | 1429 | 1944.5 | 32.66 | 6.03 | | 437 | |
| 1387 | 1902.5 | 16.13 | 4.69 | | 303 | | 1430 | 1945.5 | 32.12 | 6.34 | | 421 | |
| 1388 | 1903.5 | 23.62 | 5.34 | | 376 | | 1431 | 1946.5 | 41.06 | 8.19 | | 430 | |
| 1389 | 1904.5 | 23.59 | 5.53 | | 369 | | 1432 | 1947.5 | 42.11 | 7.98 | | 442 | |
| 1390 | 1905.5 | 26.54 | 5.88 | | 388 | | 1433 | 1948.5 | 38.57 | 7.03 | | 447 | |
| 1391 | 1906.5 | 25.16 | 5.52 | | 386 | | 1434 | 1949.5 | 37.98 | 7.87 | | 419 | |
| 1392 | 1907.5 | 29.89 | 6.11 | | 410 | | 1435 | 1950.5 | 43.28 | 7.59 | | 460 | |
| 1393 | 1908.5 | 30.67 | 6.34 | | 409 | | 1436 | 1951.5 | 49.66 | 7.54 | | 500 | |
| 1394 | 1909.5 | 29.93 | 6.50 | | 398 | | 1437 | 1952.5 | 49.00 | 7.04 | | 514 | |
| 1395 | 1910.5 | 33.64 | 6.00 | | 446 | | 1438 | 1953.5 | 41.28 | 6.23 | | 494 | |
| 1396 | 1911.5 | 30.37 | 5.48 | | 438 | | 1439 | 1954.5 | 34.38 | 5.78 | | 460 | |
| 1397 | 1912.5 | 21.12 | 5.08 | | 357 | | 1440 | 1955.5 | 34.52 | 6.19 | | 446 | |
| 1398 | 1913.5 | 20.46 | 4.87 | | 356 | | 1441 | 1956.5 | 42.18 | 7.93 | | 444 | |
| 1399 | 1914.5 | 24.78 | 5.21 | | 393 | | 1442 | 1957.5 | 45.34 | 9.11 | | 432 | |
| 1400 | 1915.5 | 31.87 | 5.82 | | 438 | | 1443 | 1958.5 | 45.23 | 8.66 | | 442 | |
| 1401 | 1916.5 | 37.97 | 6.34 | | 466 | | 1444 | 1959.5 | 48.11 | 8.21 | | 471 | |
| 1402 | 1917.5 | 35.84 | 6.90 | | 432 | | 1445 | 1960.5 | 50.59 | 9.09 | | 460 | |
| 1403 | 1918.5 | 40.92 | 6.97 | | 465 | | 1446 | 1961.5 | 37.60 | 7.18 | | 436 | |
| 1404 | 1919.5 | 40.12 | 7.09 | | 456 | | 1447 | 1962.5 | 34.92 | 6.14 | | 451 | |
| 1405 | 1920.5 | 33.86 | 6.73 | | 422 | | 1448 | 1963.5 | 33.04 | 5.91 | | 444 | |
| 1406 | 1921.5 | 30.58 | 6.24 | | 412 | | 1449 | 1964.5 | 28.63 | 5.76 | | 411 | 362 |
| 1407 | 1922.5 | 34.95 | 5.85 | | 462 | | 1450 | 1965.5 | 24.79 | 5.60 | 5.09 | 380 | 415 |
| 1408 | 1923.5 | 22.60 | 5.18 | | 371 | | 1451 | 1966.5 | 29.42 | 5.87 | 6.20 | 415 | 436 |
| 1409 | 1924.5 | 22.03 | 5.53 | | 353 | | 1452 | 1967.5 | 32.25 | 6.86 | 6.38 | 406 | 426 |
| 1410 | 1925.5 | 26.97 | 6.00 | | 388 | | 1453 | 1968.5 | 35.34 | 6.42 | 6.24 | 444 | 464 |
| 1411 | 1926.5 | 36.03 | 6.95 | | 432 | | 1454 | 1969.5 | 31.28 | 6.40 | 6.09 | 412 | 418 |
| 1412 | 1927.5 | 29.99 | 6.49 | | 399 | | 1455 | 1970.5 | 31.93 | 6.59 | 6.44 | 412 | 420 |
| 1413 | 1928.5 | 31.74 | 6.43 | | 415 | | 1456 | 1971.5 | 32.48 | 6.26 | 5.97 | 427 | 439 |
| 1414 | 1929.5 | 34.93 | 6.52 | | 437 | | 1457 | 1972.5 | 33.80 | 6.40 | 6.43 | 433 | 403 |
| 1415 | 1930.5 | 47.20 | 6.77 | | 513 | | 1458 | 1973.5 | 41.24 | 6.31 | 6.21 | 491 | 485 |
| 1416 | 1931.5 | 31.56 | 5.72 | | 439 | | 1459 | 1974.5 | 46.78 | 6.40 | 6.66 | 525 | 531 |

| | | | | | | |
|---|---|---|---|---|---|---|
| 1460 | 1975.5 | 38.38 | 5.93 | 5.90 | 485 | 480 |
| 1461 | 1976.5 | 35.66 | 6.04 | 5.42 | 460 | 451 |
| 1462 | 1977.5 | 34.82 | 6.28 | 5.96 | 445 | 418 |
| 1463 | 1978.5 | 40.10 | 7.29 | 7.27 | 449 | 429 |
| 1464 | 1979.5 | 37.82 | 7.24 | 7.59 | 435 | 418 |
| 1465 | 1980.5 | 30.65 | 6.71 | 6.96 | 398 | 389 |
| 1466 | 1981.5 | 39.91 | 7.90 | 7.87 | 430 | 424 |
| 1467 | 1982.5 | 51.99 | 8.46 | 8.93 | 485 | 469 |
| 1468 | 1983.5 | 47.95 | 7.07 | 8.01 | 506 | 477 |
| 1469 | 1984.5 | 45.87 | 6.81 | 7.81 | 503 | 471 |
| 1470 | 1985.5 | 38.72 | 6.19 | 5.95 | 478 | 472 |
| 1471 | 1986.5 | 34.86 | 6.14 | 5.74 | 450 | 459 |
| 1472 | 1987.5 | 32.97 | 5.93 | 6.26 | 442 | 428 |
| 1473 | 1988.5 | 35.47 | 6.62 | 7.31 | 438 | 428 |
| 1474 | 1989.5 | 46.07 | 9.12 | 8.20 | 436 | 460 |
| 1475 | 1990.5 | 41.09 | 7.51 | 7.39 | 449 | 434 |
| 1476 | 1991.5 | 52.63 | 8.52 | 9.35 | 486 | 467 |
| 1477 | 1992.5 | 43.68 | 7.53 | 8.26 | 465 | 440 |
| 1478 | 1993.5 | 40.60 | 6.68 | 6.52 | 473 | 451 |
| 1479 | 1994.5 | 44.60 | 6.30 | 6.35 | 514 | 514 |
| 1480 | 1995.5 | 36.24 | 6.30 | 5.72 | 455 | 426 |
| 1481 | 1996.5 | 30.56 | 5.56 | 5.18 | 436 | 421 |
| 1482 | 1997.5 | 29.32 | 5.93 | 5.54 | 411 | 381 |
| 1483 | 1998.5 | 35.62 | 6.78 | 6.89 | 434 | 409 |
| 1484 | 1999.5 | 36.54 | 6.56 | 6.87 | 448 | 439 |
| 1485 | 2000.5 | 39.87 | 7.80 | 7.14 | 433 | 446 |
| 1486 | 2001.5 | 35.86 | 7.84 | 6.80 | 405 | 427 |
| 1487 | 2002.5 | 36.93 | 6.97 | 7.69 | 437 | 439 |
| 1488 | 2003.5 | 54.28 | 7.53 | 7.54 | 526 | 545 |
| 1489 | 2004.5 | 37.90 | 6.90 | 6.55 | 447 | 453 |
| 1490 | 2005.5 | 37.63 | 6.50 | 6.23 | 458 | 473 |
| 1491 | *2006.5* | *27.84* | *5.2* | *4.96* | *416* | *428* |
| 1492 | *2007.2* | *27.4* | *5.0* | *4.7* | *430* | *440* |

**Table 2.** Yearly values of composite *IHV*, *B* derived from the *IDV*-index, *V* calculated using eq.(11), and *B* and *V* observed by spacecraft. *B* in nT and *V* in km/s. Values for 2006 and 2007 are preliminary only, based on incomplete data.

| Long. | Lat | Name | From-To | Name | From-To | Name | From-To |
|---|---|---|---|---|---|---|---|
| 0E | 41N | Ebro | 1910-… | | | | |
| 3E | 48N | Saint Maur | 1883-1900 | Val Joyeux | 1901–1936 | Chambon la Forêt | 1936-… |
| 5E | 50N | Uccle | 1896-1919 | Manhay | 1936–1971 | Dourbes | 1955-… |
| 6E | 52N | Utrecht | 1891-1896 | De Bilt | 1899–1938 | Witteveen | 1938-1988 |
| 8E | 54N | Wilhelmshafen | 1883-1911 | Wingst | 1939-… | | |
| 11E | 60N | Oslo | 1843-1930 | | | | |
| 13E | 52N | Potsdam | 1890-1907 | Seddin | 1908–1931 | Niemegk | 1932-… |
| 13E | 57N | Copenhagen | 1891-1908 | Rude Skov | 1907-1978 | Brorfelde | 1978-… |
| 25E | 60N | Helsinki | 1844-1897 | Nurmijärvi | 1953-… | | |
| 31E | 30N | Helwan | 1903-1959 | Missalat | 1960-… | | |
| 31E | 60N | St. Petersburg | 1869-1877 | Slutsk | 1878–1941 | Voeikovo | 1947-… |
| 45E | 42N | Tiflis | 1879-1905 | Karsani | 1905–1934 | Dusheti | 1938-… |
| 48E | 19S | Antananarivo | 1890-… | | | | |
| 49E | 56N | Kazan | 1892-1913 | | | | |
| 58E | 20S | Mauritius | 1892-1965 | | | | |
| 61E | 57N | Sverdlovsk | 1887-1978 | Vysokaya-Dubrava | 1932-… | | |
| 73E | 19N | Colaba | 1846-1905 | Alibag | 1904-… | | |
| 77E | 10N | Kodaikanal | 1902-… | | | | |
| 107E | 6S | Batavia | 1867-1944 | Kuyper | 1950–1962 | Tangerang | 1964-… |
| 114E | 22N | Hong Kong | 1884-1928 | Au Tau | 1928–1939 | Hong Kong | 1972-… |
| 121E | 15N | Manila | 1891-1904 | Antipolo | 1910–1938 | Muntinlupa | 1951-… |
| 121E | 31N | Zi-Ka-Wei | 1875-1907 | Lukiapang | 1908–1933 | Zo-Se | 1933-1974 |
| 145E | 38S | Melbourne | 1865-1921 | Toolangui | 1922–1980 | Canberra | 1980-… |
| 173E | 43S | Amberley | 1929-1977 | Lauder | 1977–1978 | Eyrewell | 1979-… |
| 188E | 14S | Apia | 1905-… | | | | |
| 261E | 20N | Teoloyucan | 1923-… | | | | |
| 281E | 44N | Agincourt | 1881-1969 | Ottawa | 1968-… | | |
| 294E | 22S | La Quiaca | 1920-… | | | | |
| 296E | 32S | Pilar | 1905-… | | | | |
| 316E | 22S | Vassouras | 1915-… | | | | |
| 334E | 38N | Sao Miguel | 1911-… | | | | |
| 352E | 40N | Coimbra | 1866-… | | | | |
| 356E | 36N | San Fernando | 1891-… | | | | |
| 358E | 54N | Stonyhurst | 1865–1967 | | | | |
| 360E | 51N | Greenwich | 1846-1925 | Abinger | 1925–1957 | Hartland | 1957-… |
| 360E | 51N | Kew | 1858-1924 | | | | |

**Table 3.** Geomagnetic observatories with long series of data that may be useful for constructing *IHV*-indices. If a station stopped observing, the next column(s) may give the replacement station(s) (if any). For many stations there are data even earlier than given here, *e.g.* Paris and Munich. The coordinates given in the first column are geographic longitude and latitude.

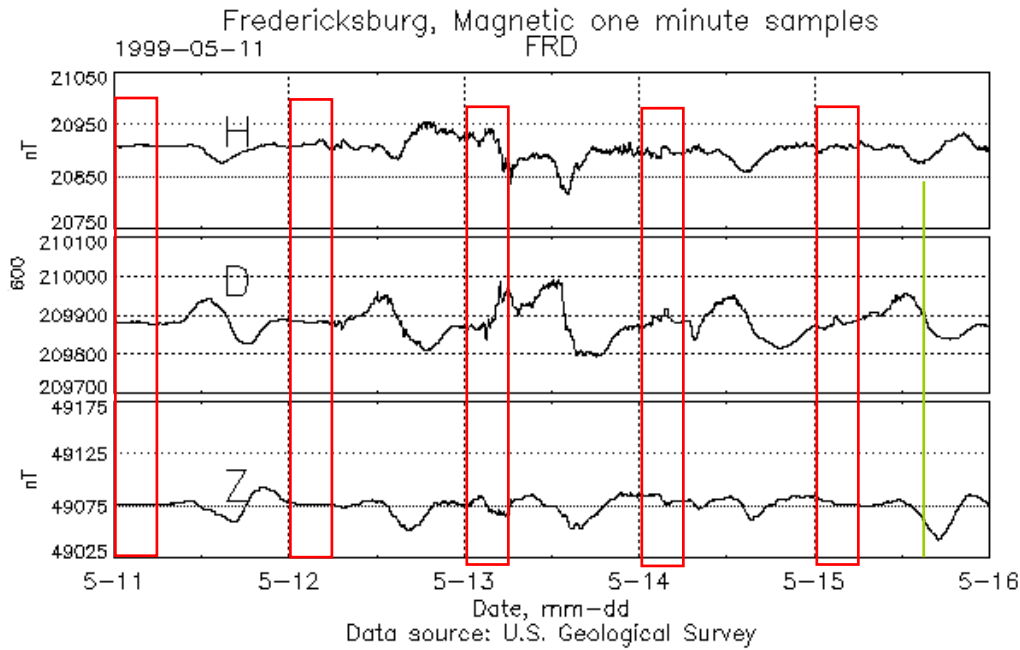

**Figure 1.** Variation of the geomagnetic elements at Fredericksburg May 11-15, 1999 (UT). The "effective" noon is marked with a green line on May 15. The red boxes indicate the six hours around midnight where the regular variation is absent or minimal. These intervals are used to define the *IHV*-index. May 11 is a good example of a day with very little activity. It is, in fact, the famous day where "the solar wind disappeared" [*e.g. Jordanova et al.*, 2001]. The solar wind momentum flux was only 1% of its usual value and the magnetosphere diameter was five times larger than normal. The interplanetary magnetic field was not affected and had its usual properties. The variability of $S_R$ is clearly seen by comparing May 11 and May 15.

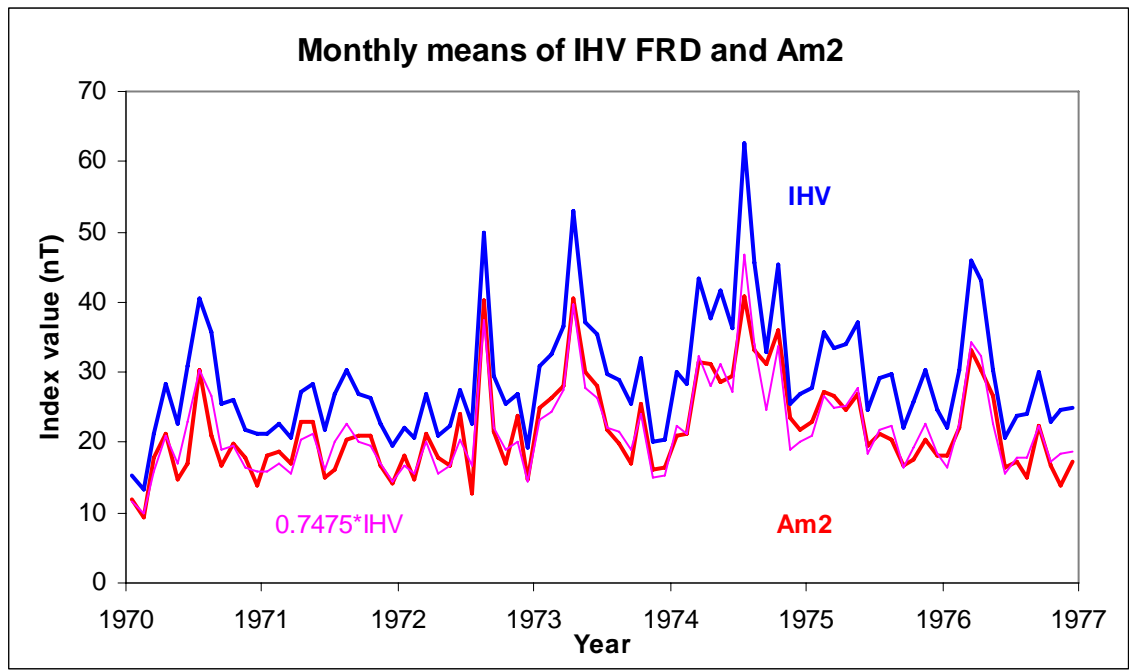

**Figure 2.** Comparison of monthly means of the "raw" *IHV*-index (blue) calculated for the H-component at Fredericksburg and the *Am2*-index (red) for the interval 1970-76. The year-labels on the abscissa mark the beginning of each year. The thin pink curve shows *IHV* scaled down by 0.7475 for direct comparison with *Am2*.

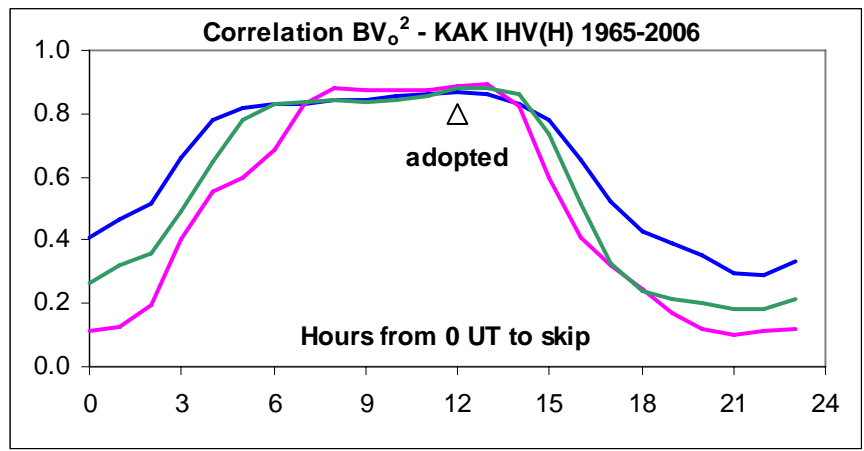

**Figure 3.** Correlation between yearly averages of *IHV* calculated for KAK for the interval 1965-2006 and the quantity $BV^2$ (see section 10.3) as a function of the number of hours from $0^h$ UT to skip before calculating *IHV*. Blue curve is for the H-component, green for the Z-component, and pink for the D-component. The triangle shows the correlation for the number of "skip hours" adopted for this station (12 in this case).

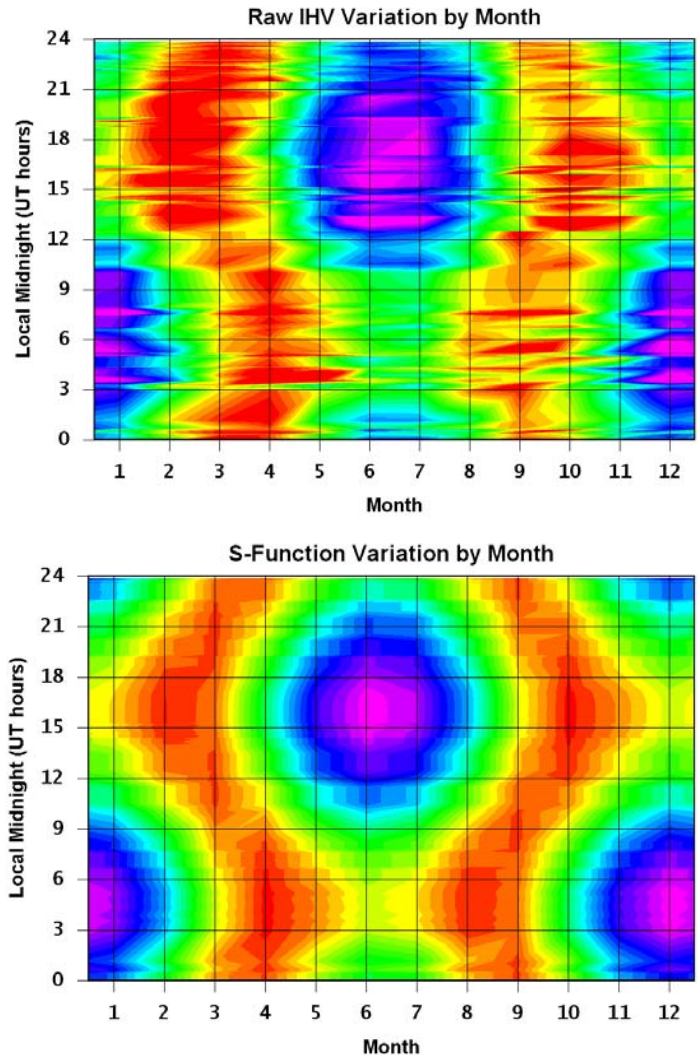

**Figure 4.** Variation of the *S*-function (bottom panel) and of "raw" *IHV* (top panel) with month of year and Universal Time calculated for all the stations in Table 1 for *all* data available for each station. The *IHV* values for a given station were assigned to the Universal Time of local midnight. All values were divided by the average values for each station. The color coding over the ~40% variation is chosen such that purple to red represents low to high values.

1572

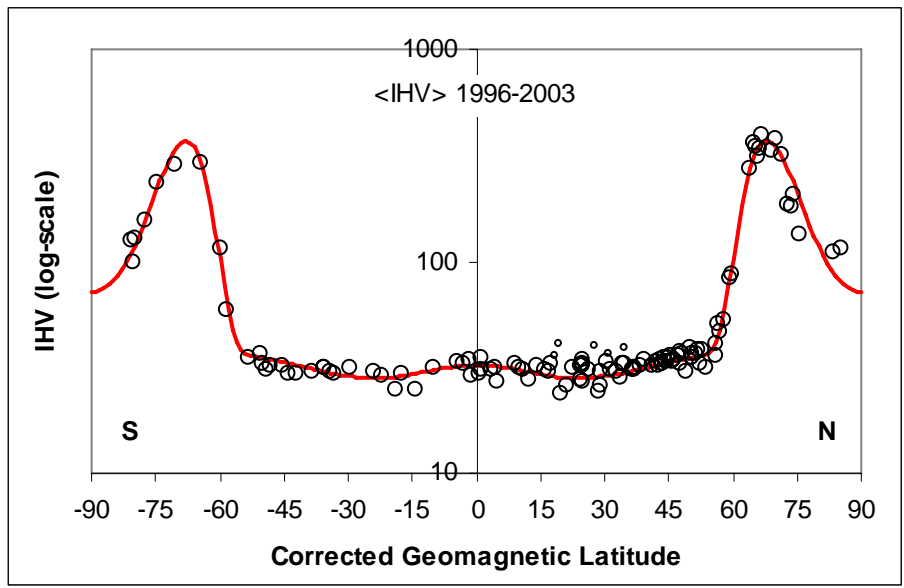

**Figure 5.** Variation of *IHV* with corrected geomagnetic latitude. Average *IHV* over the interval 1996-2003 for each station with data in that interval are plotted. A few "outliers" (SIL, KRC, QSB, GLM, and KSH) are shown with small circles. Local induction effects may be responsible for these stations having about 25% higher *IHV*. The red curve shows a model fit to the larger circles as described in the text.

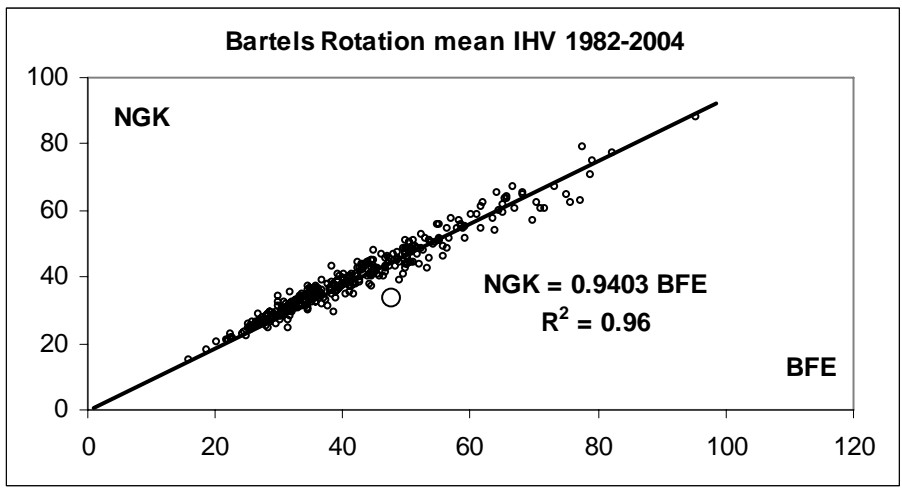

**Figure 6.** Bartels rotation means of *IHV* for BFE versus NGK for 1982-2004. The scale factor is derived as the slope of the regression line constrained to go through the origin. A single outlier marked with a large circle is not included in the fit.

1622

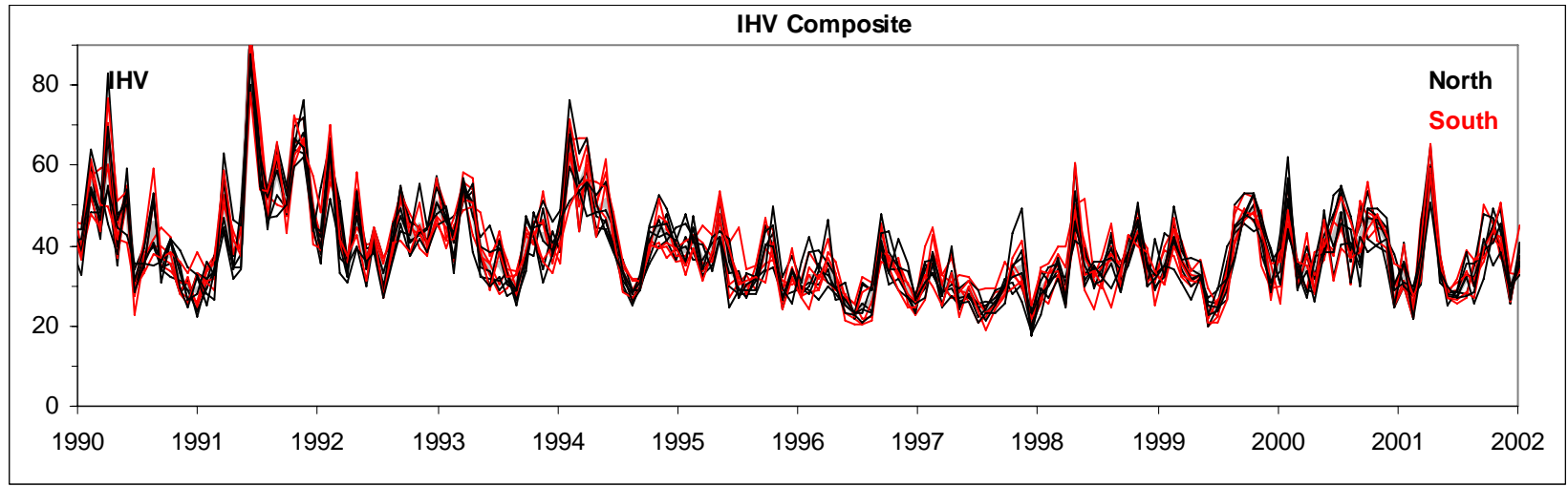

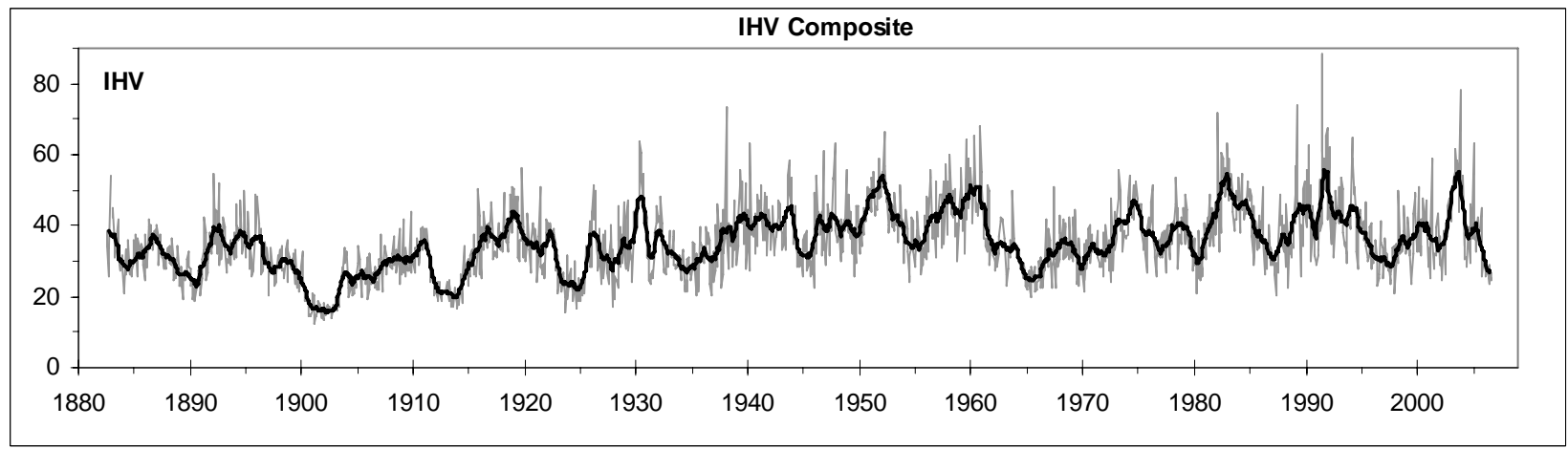

1623
1624

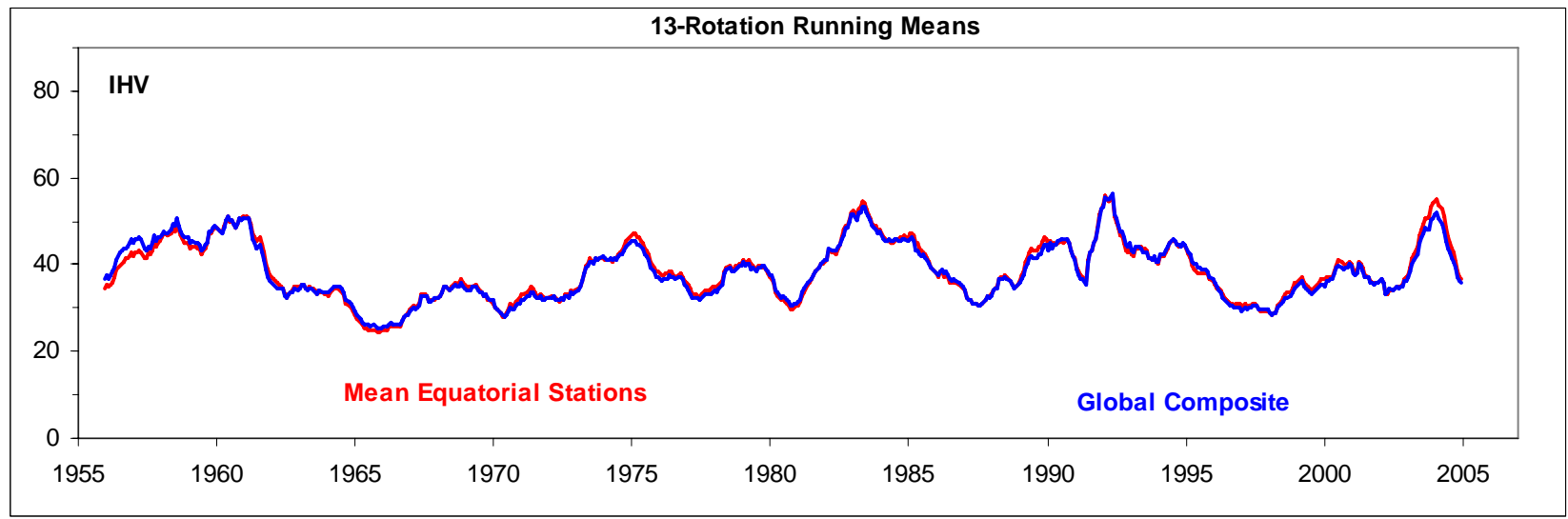

Figure 7. (Upper) Plot of a portion (years 1990-2001) of all the individual data series that went into the composite series. Northern sectors are shown in black while southern sectors are shown in red. (Middle) Plot of the full series for years 1883-2006 (grey curve) overlain by its 13-rotation running mean (heavy black curve). The curve before 1890 is based on preliminary data from BTV and WLH. (Lower) Plot of 13-rotation running means of the composite *IHV* (blue) and *IHV* derived from Equatorial stations (red).

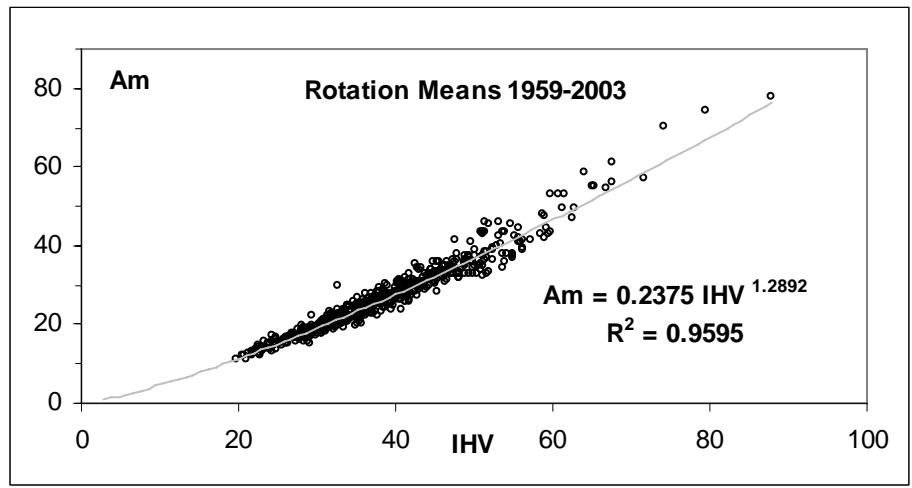

**Figure 8.** Relationship between Bartels rotation averages of the *Am*-index (freed for dipole tilt effect) and composite *IHV* for the interval 1959-2003.

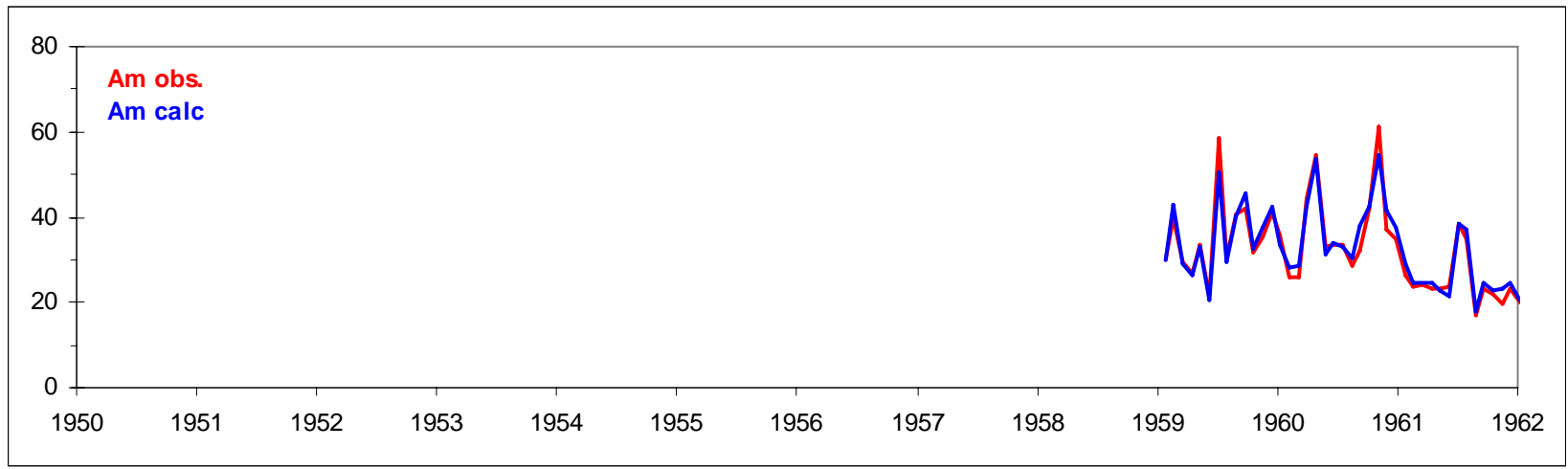
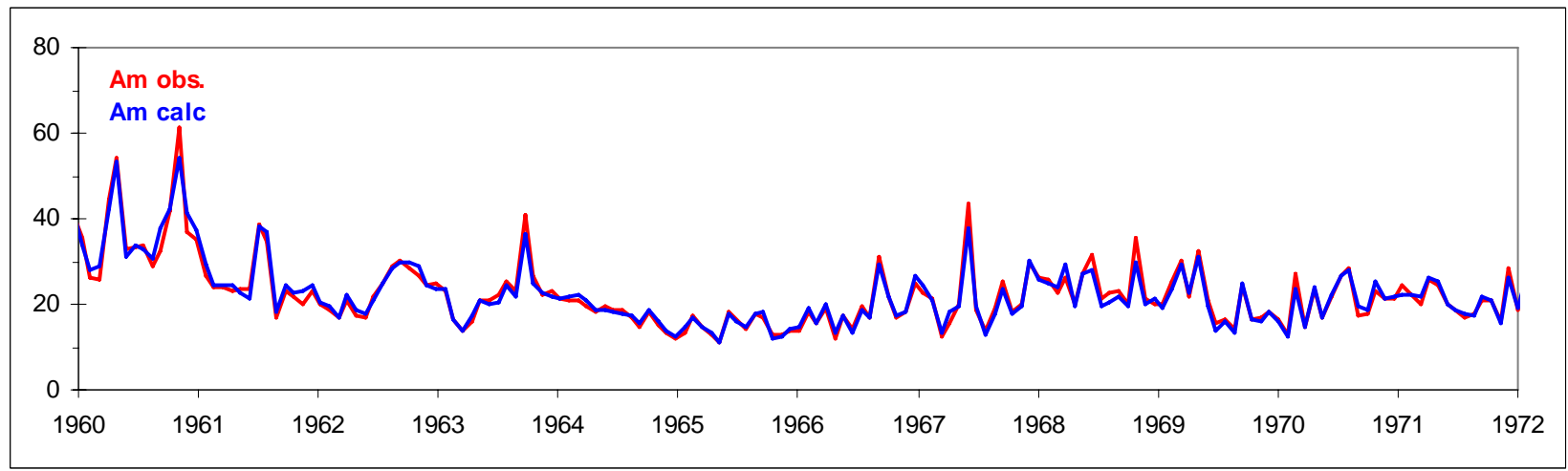

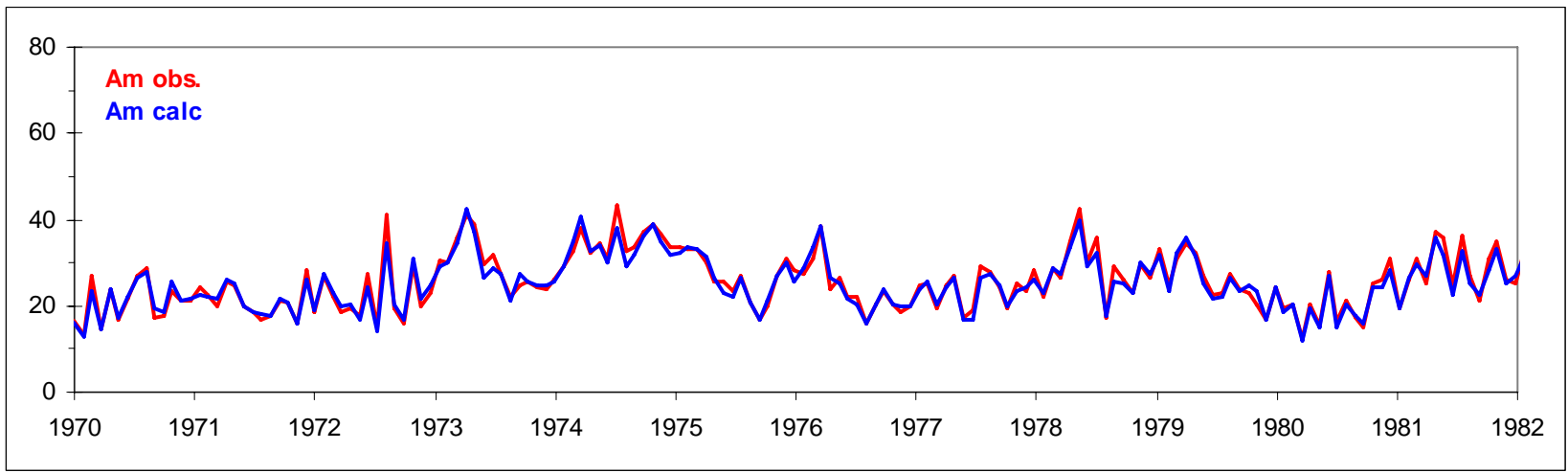
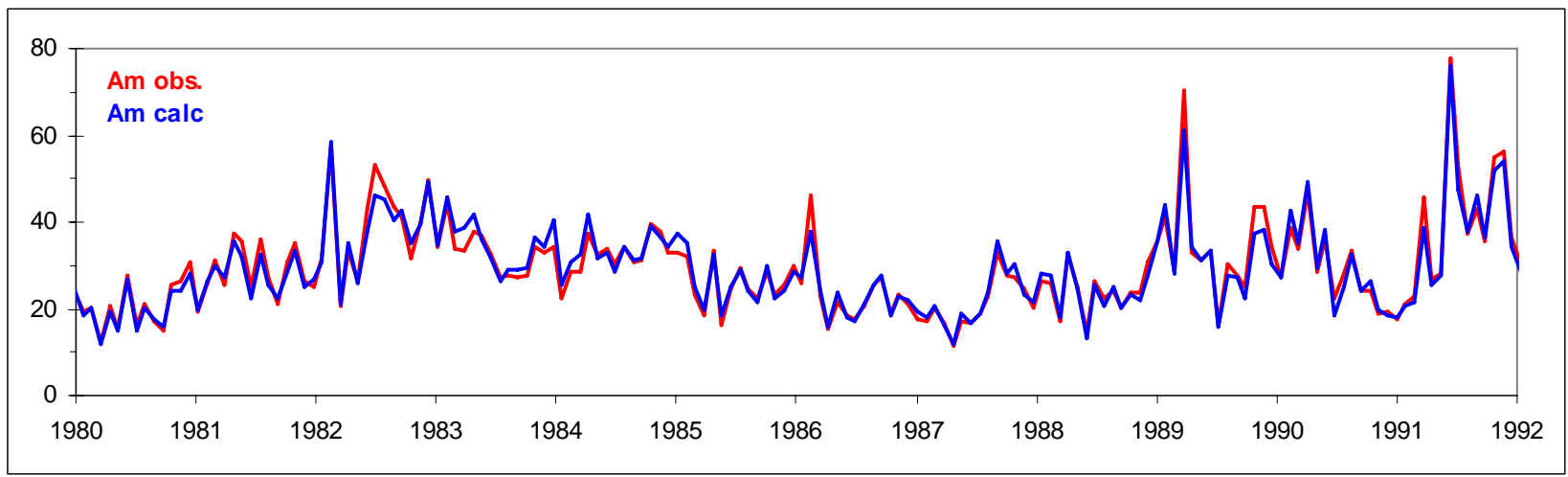

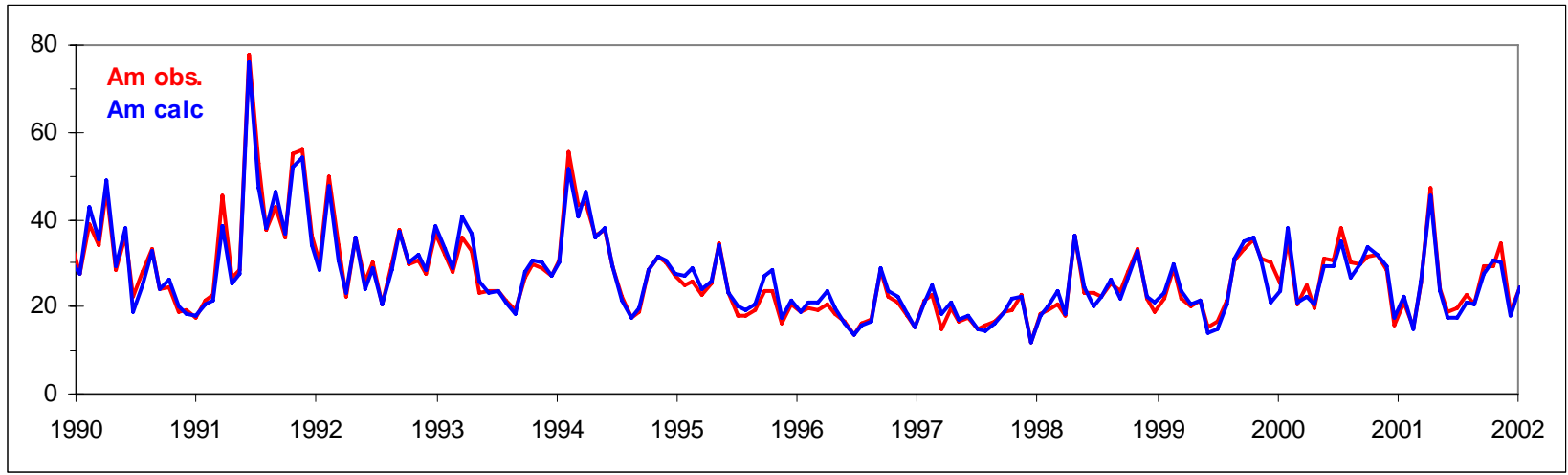
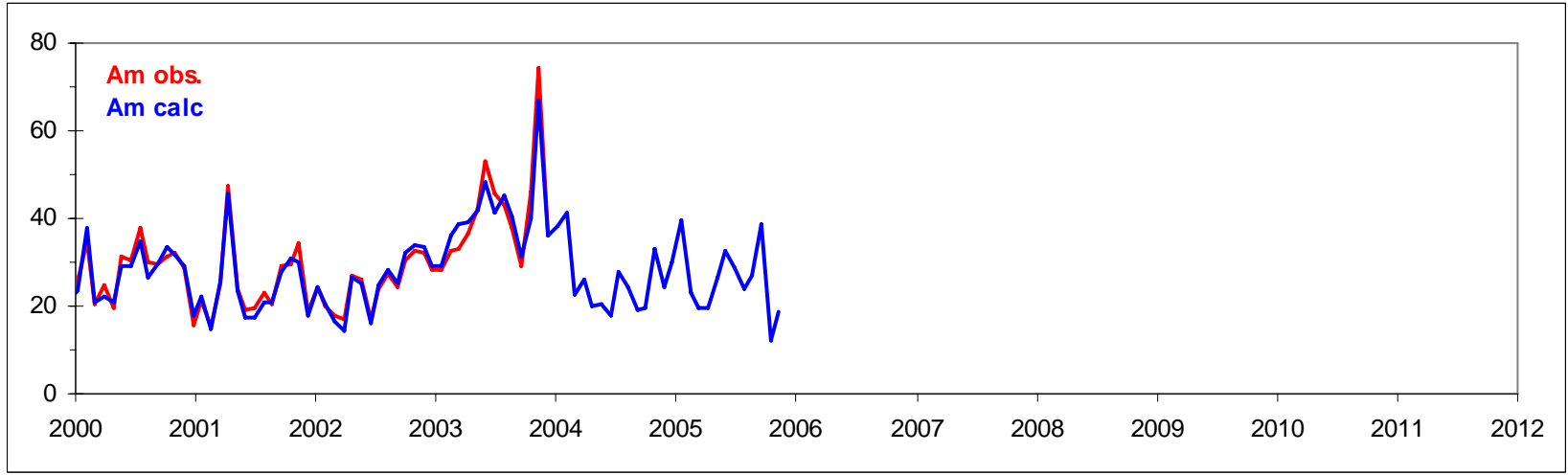

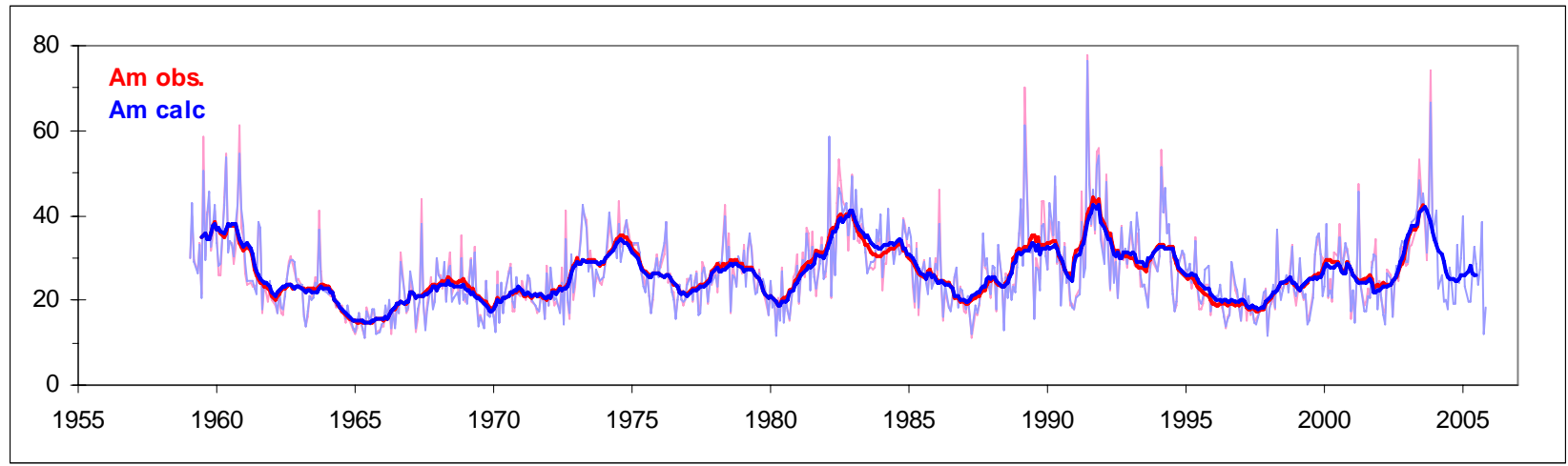

**Figure 9.** (Upper panels) Bartels rotation averages of proxy values of *Am* calculated using eq.(4) (blue curve) and observed (red curve). Both datasets have been freed from the effect of the dipole tilt (section 3.2). The bottom panel shows the entire datasets overlain by their 13-rotation running means.

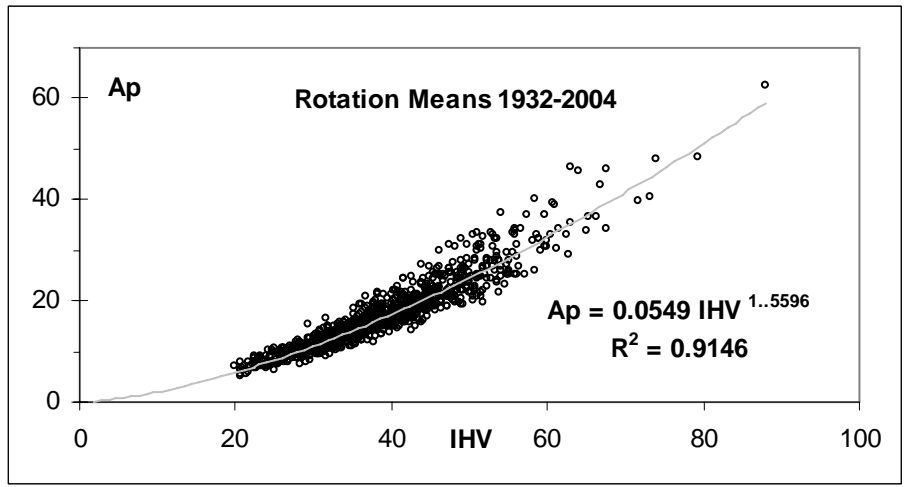

**Figure 10.** Relationship between Bartels rotation means of *Ap* (freed for dipole tilt effect) and composite *IHV* for the interval 1932-2004.

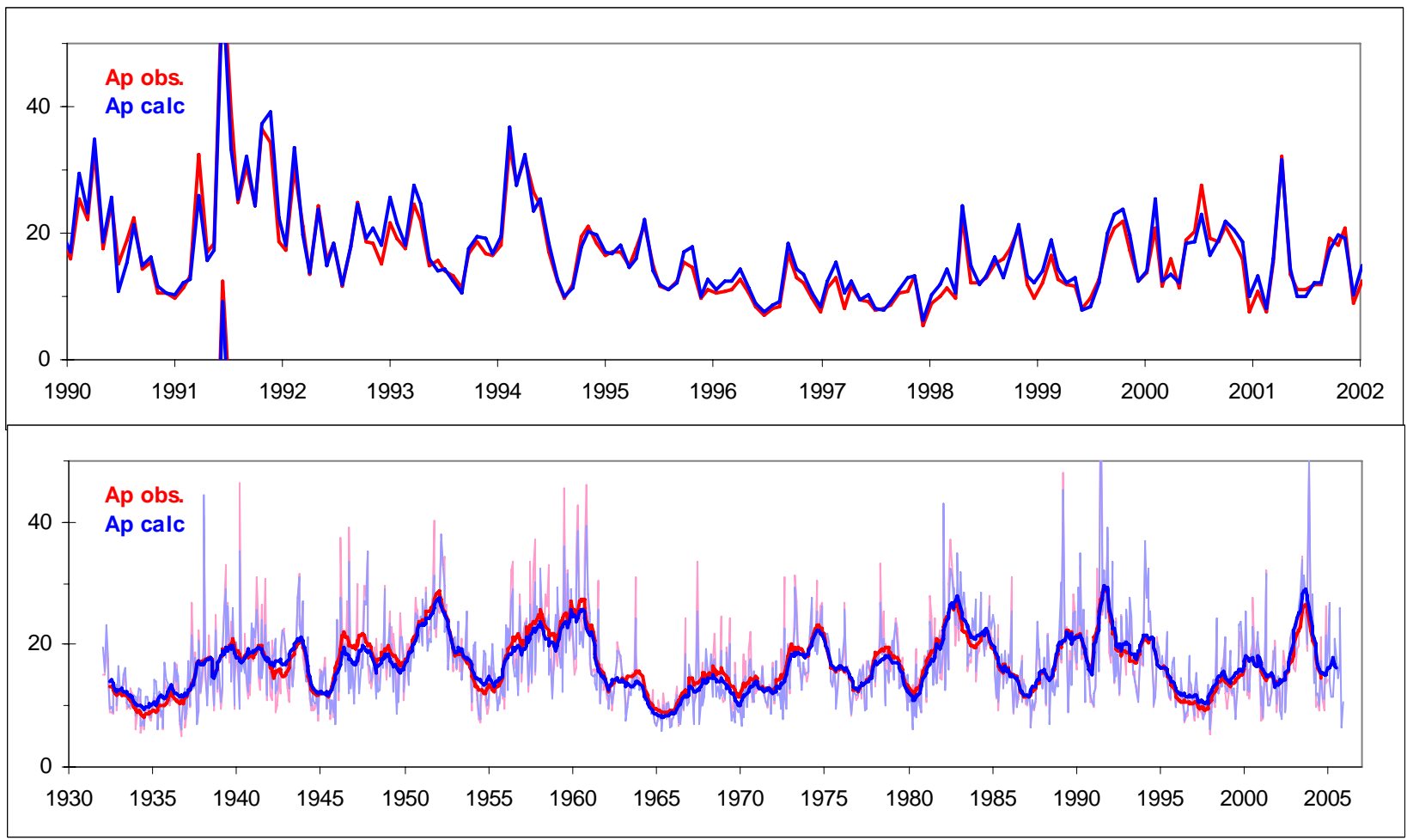

**Figure 11.** (Upper) Sample Bartels rotation averages of proxy values of *Ap* calculated using eq.(5) (blue curve) and observed (red curve). Both datasets have been freed from the effect of the dipole tilt (section 3.2). (Lower) Shows the entire datasets overlain by their 13-rotation running means.

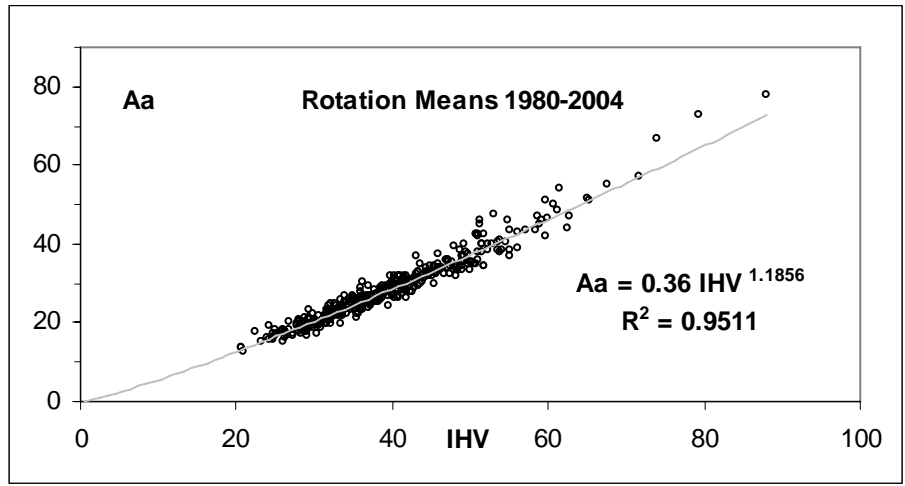

**Figure 12.** Relationship between Bartels rotation means of *Aa* (freed for dipole tilt effect) and composite *IHV* for the interval 1980-2004.

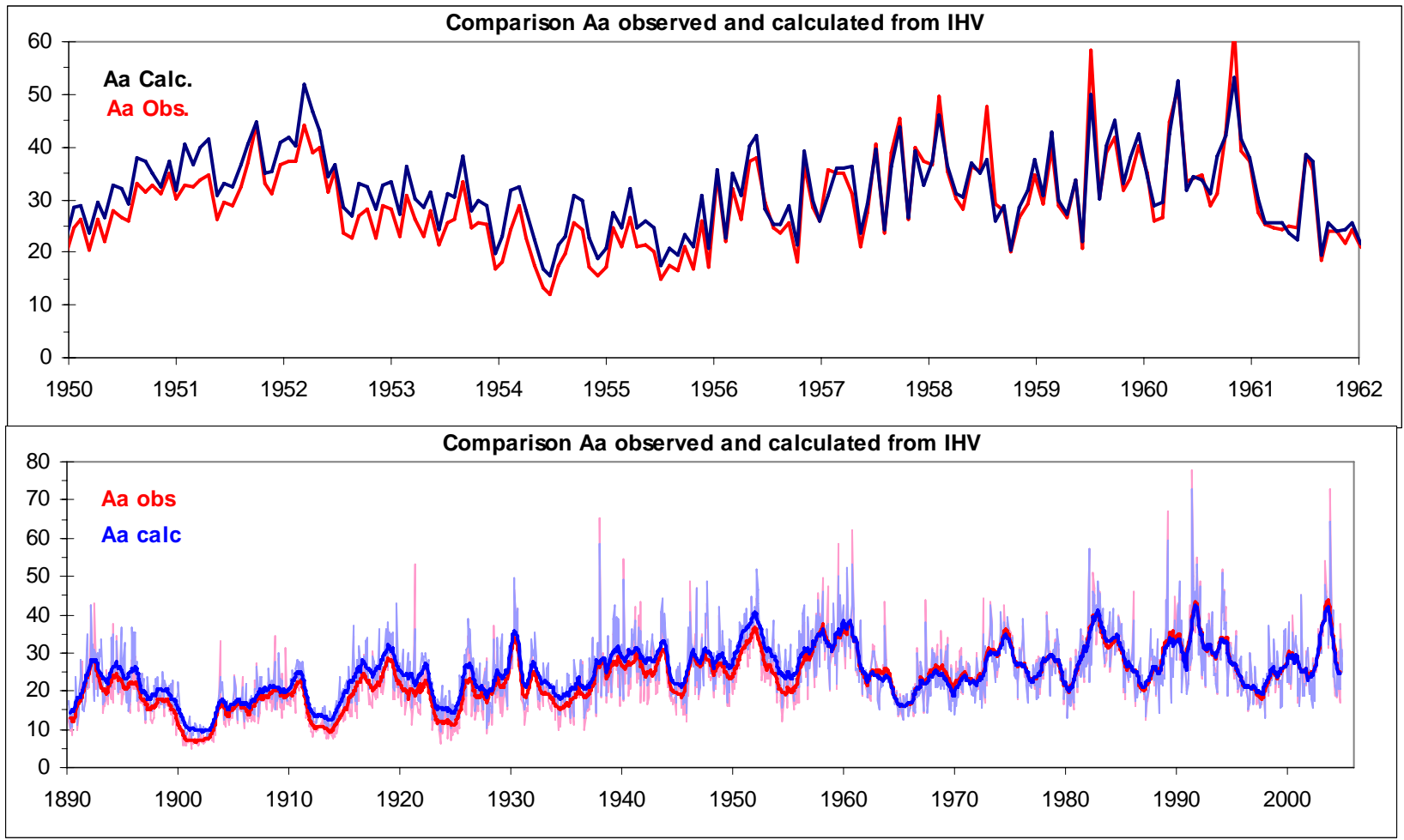

**Figure 13.** (Upper) Sample Bartels rotation averages of proxy values of *Aa* calculated using eq.(6) (blue curve) and observed (red curve). Both datasets have been freed from the effect of the dipole tilt (section 3.2). (Lower) Shows the entire datasets overlain by their 13-rotation running means.

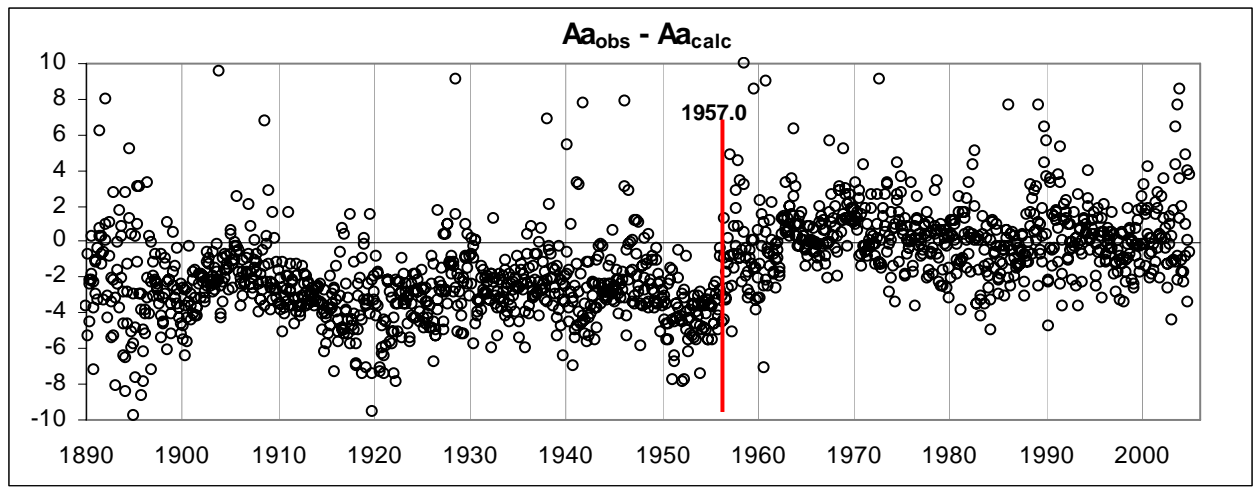

**Figure 14.** Difference between observed and calculated values of Bartels rotation means of *Aa* showing the upward jump in 1957.

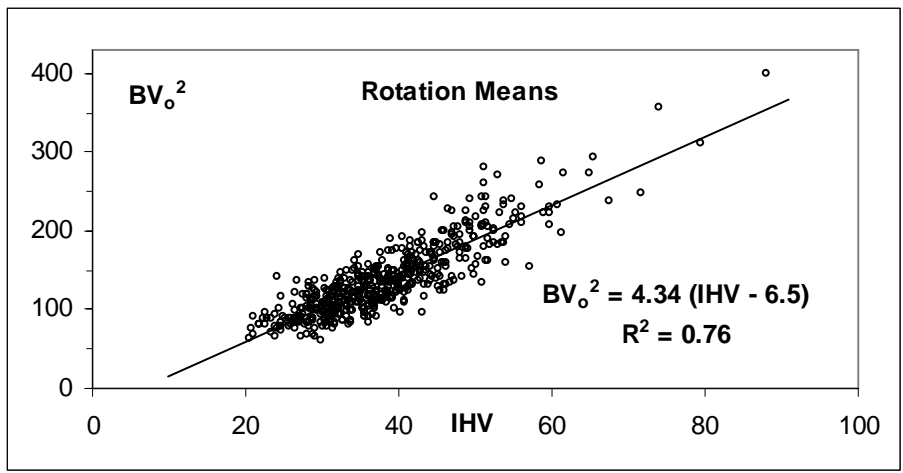

**Figure 14.** Relationship between Bartels rotation means of $BV_o^2$ and composite $IHV$ for the interval 1965-2005.

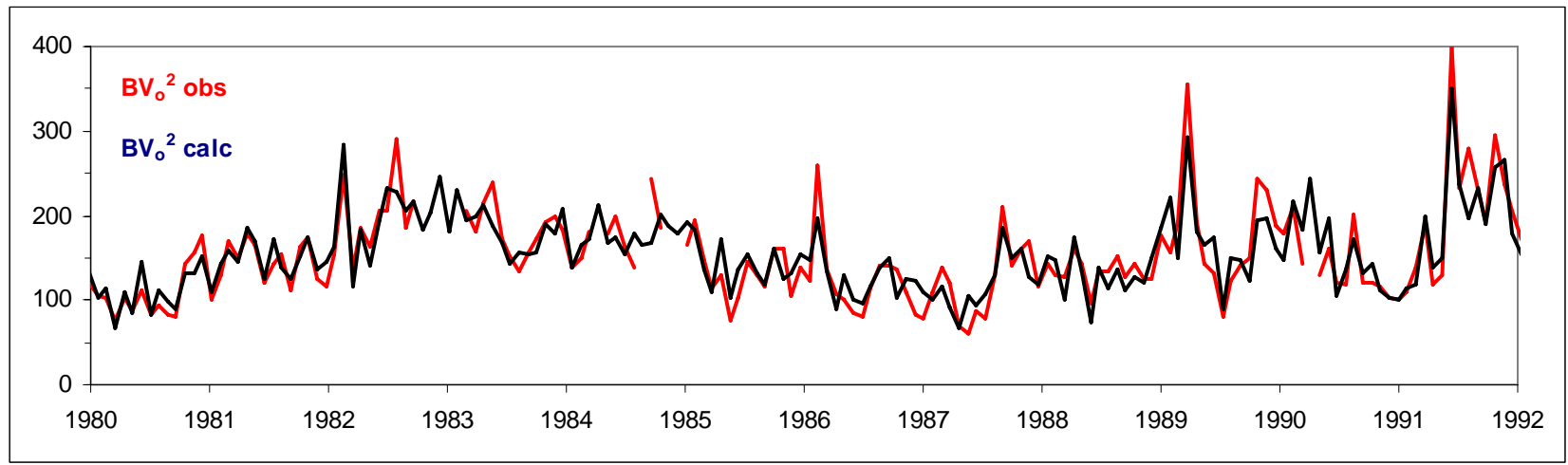

**Figure 16.** Sample Bartels rotation averages of proxy values of $BV_o^2$ calculated using eq.(7) (blue curve) and observed (red curve).

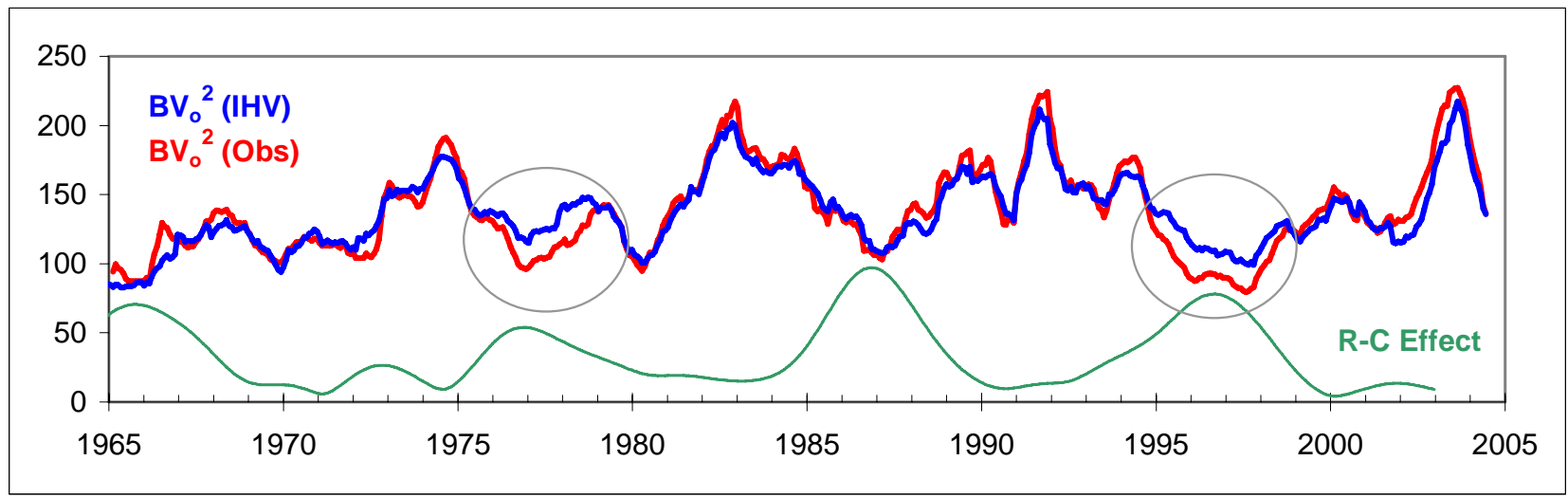

**Figure 17.** 13-rotation running means of $BV_o^2$, calculated (blue curve) and observed (red curve). Areas of consistent disagreement are marked by ovals. These occur every other time when the Rosenberg-Coleman effect is large (amplitude on arbitrary scale given by green curve).

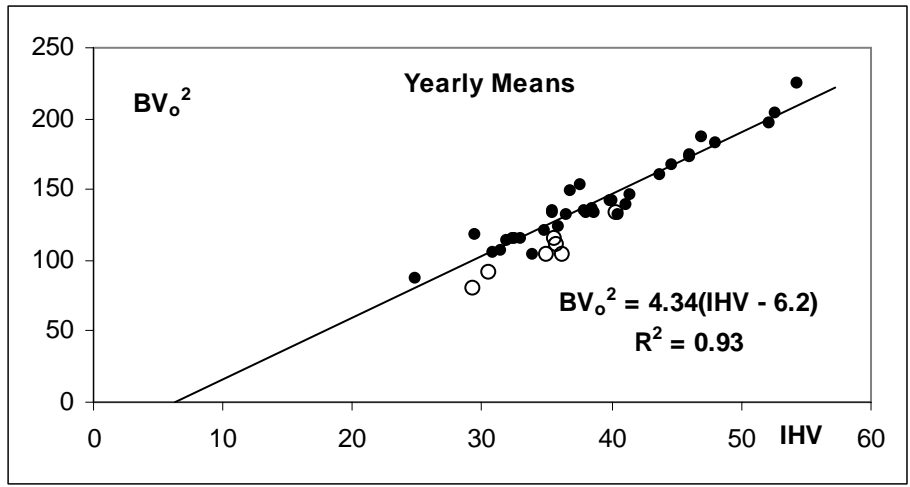

**Figure 18.** Relationship between yearly means of $BV_o^2$ and composite *IHV* for the interval 1965-2005. Years affected by the 22-year cycle are shown as open circles and are not included in the fit.

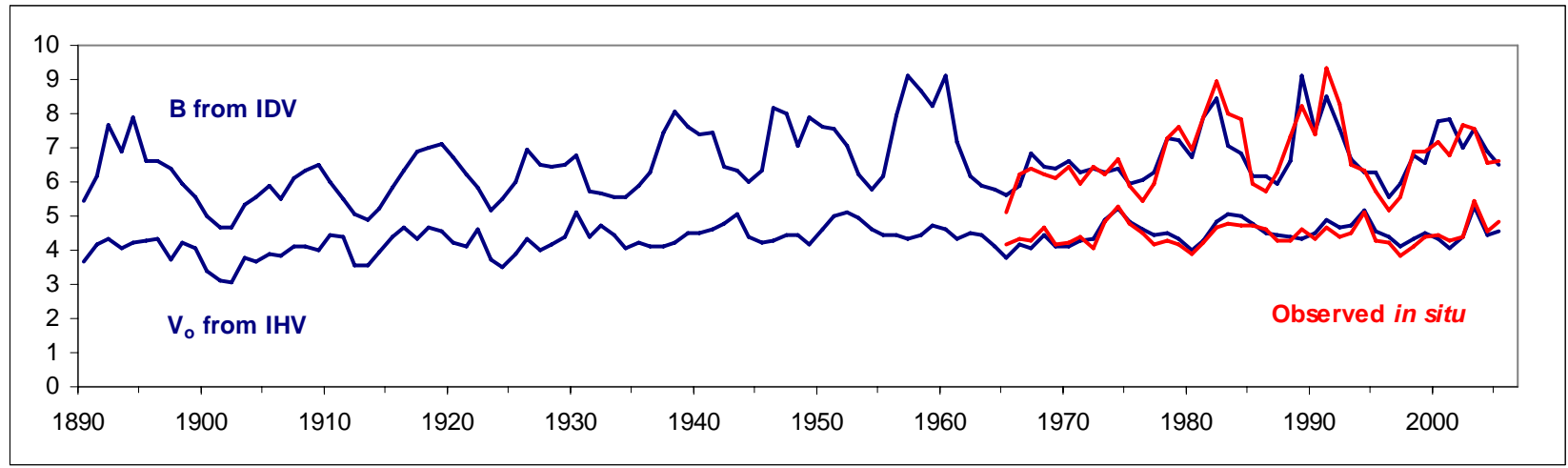

**Figure 19.** Yearly values of *B* (nT) derived from the *IDV*-index (eq.(10), upper blue curve) and *V* calculated using eq.(11) (lower blue curve). *V* is plotted as $V_o = V/100$ km/s. *B* and *V* observed in Space are shown in red.

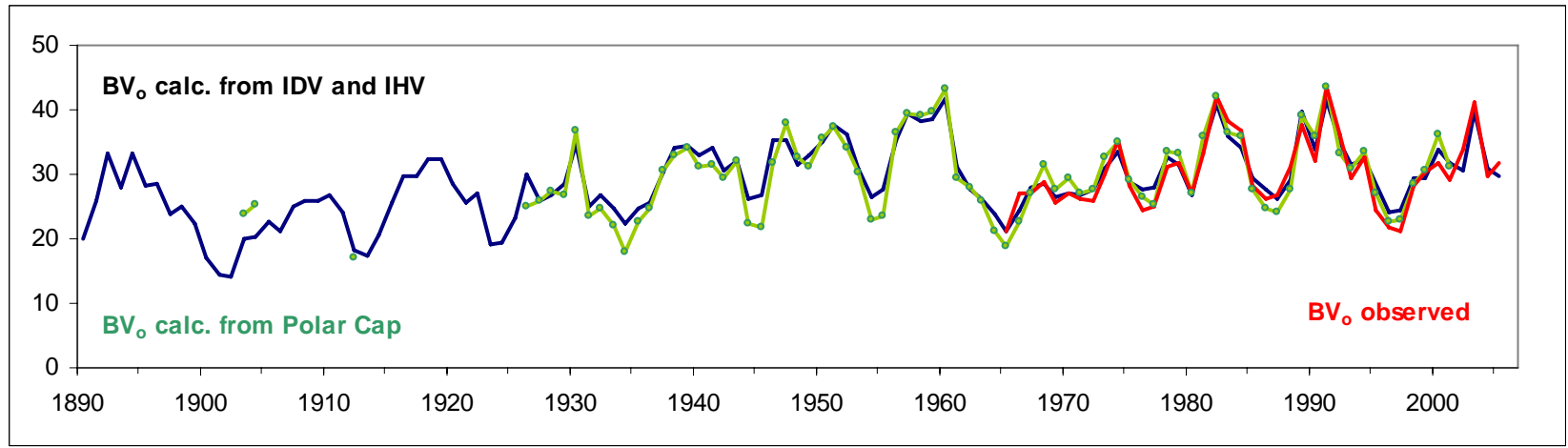

**Figure 20.** Yearly values of $BV_o$ (blue curve) calculated from $B$ derived from the *IDV*-index (eq.(10)) and $V$ derived from the IHV-index (eq.(11)) compared to $BV_o$ (green curve) calculated from the range of diurnal variation of the horizontal component in the polar caps. $BV_o$ calculated from $B$ and $V$ observed in Space is shown in red.

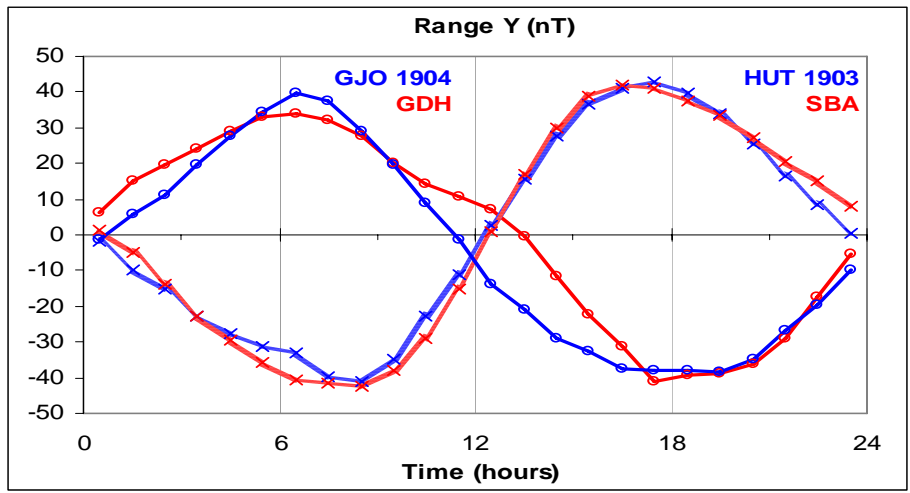

**Figure 21.** The diurnal variation of the Y-component of the geomagnetic field at SBA (red, crosses), HUT (blue crosses; for 1902.5-1903.5), GJO (blue circles; for 1904), and GDH (red circles) all in a local coordinate system where the X-axis coincides with the average direction of the H-component. The curves have been shifted in time to have the same phase in each hemisphere, but out of phase between hemispheres. This is simply a presentation device to avoid having the curves crowd on top of each other. For SBA and GDH, modern data was used for years with approximately the same sunspot activity as during 1903-1904 as described in the text.

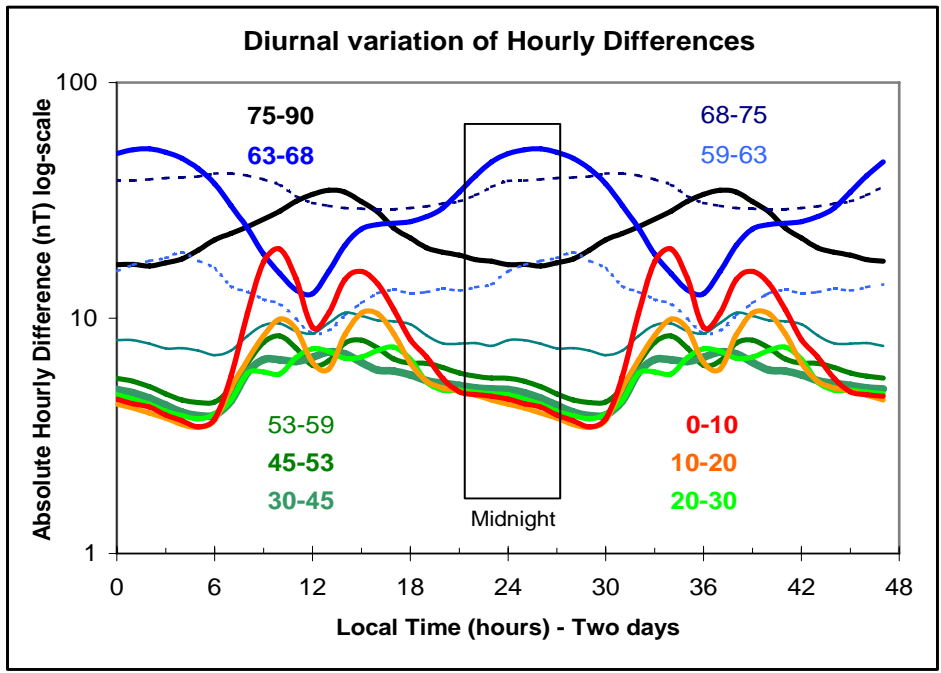

**Figure 22.** Diurnal variation of unsigned hourly differences (between one hour and the next) for the H-component as a function of local time shown for corrected geomagnetic latitude bands for all available stations during 1996-2003 (color-coded from red at the equator through green at midlatitudes to blue and black in the polar regions). A band contains all stations from both hemispheres described in section 4.5. The time extends over two days to position the six-hour midnight interval used for *IHV* in the middle of the Figure.

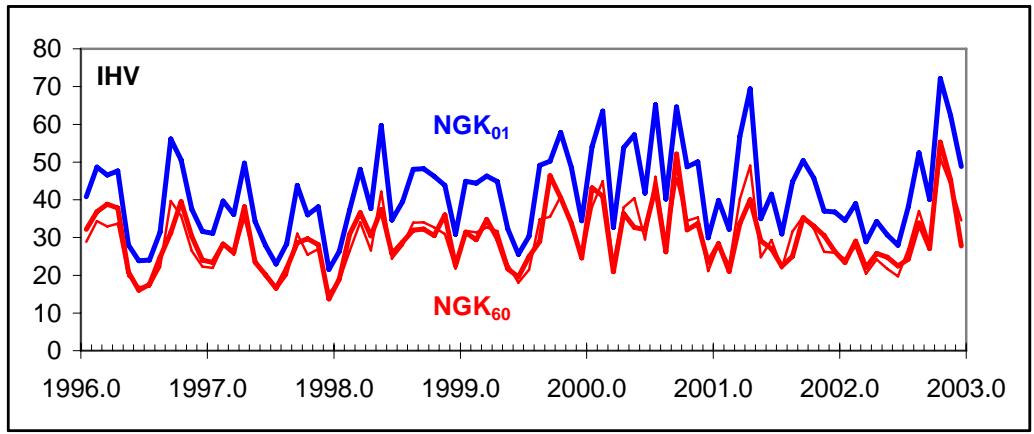

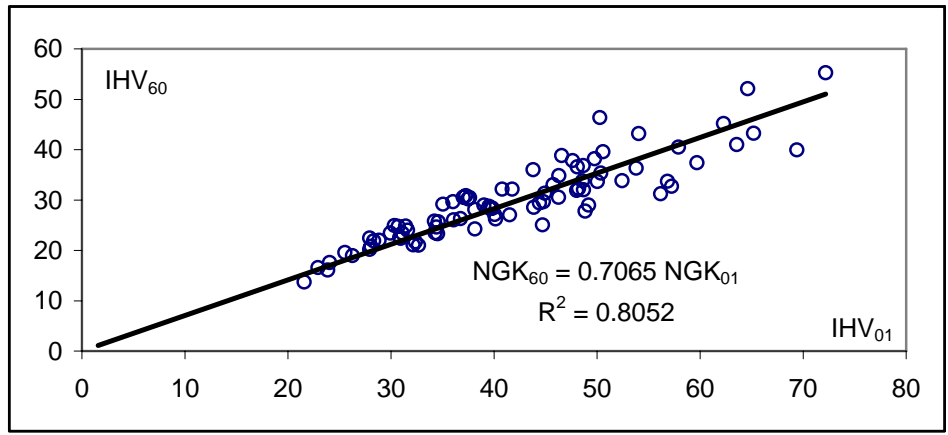

**Figure 23.** (Upper) Monthly means of *IHV* for Niemegk (NGK) 1996-2002. The heavy red curve shows *IHV* calculated from true hourly means (calculated as the mean of 60 one-minute values of the H-component). The blue curve shows *IHV* calculated from a single one-minute value taken each hour on the hour. The thin red curve shows the blue curve scaled down by the coefficient determined by the linear regression shown in the lower panel. (Lower) Correlation between the monthly means of *IHV* shown in the upper panel calculated from hourly means ($IHV_{60}$) versus calculated from hourly values (one-minute averages taken once an hour, $IHV_{01}$).

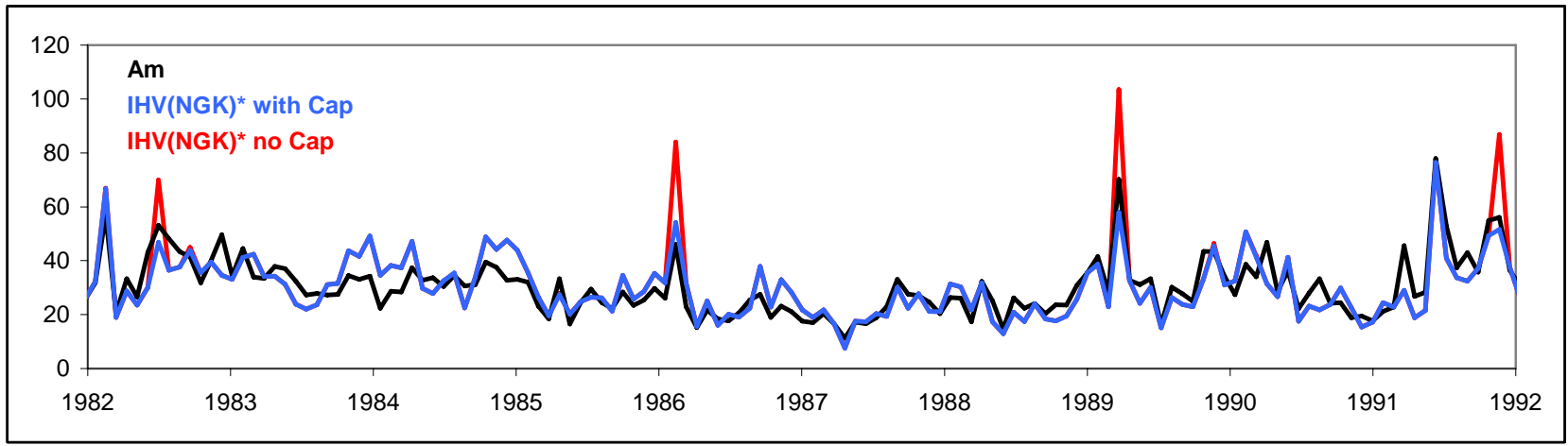

**Figure 24**. Rotation averages of *Am* (black curve) compared to *IHV* from NGK (blue curve) [scaled to *Am* using eq. (4)] derived using the cap. The red curve (almost always hidden behind the blue curve) shows what *IHV* would have been without the cap.

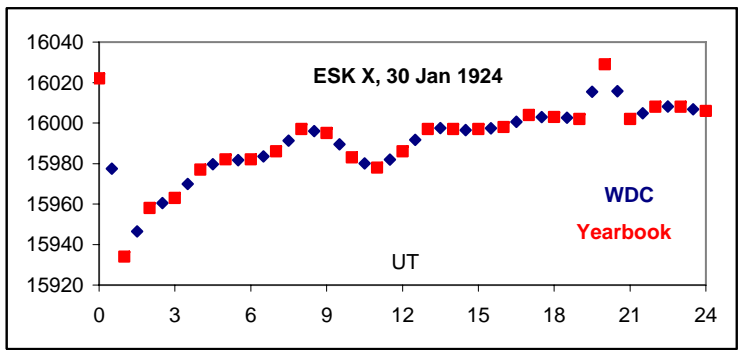

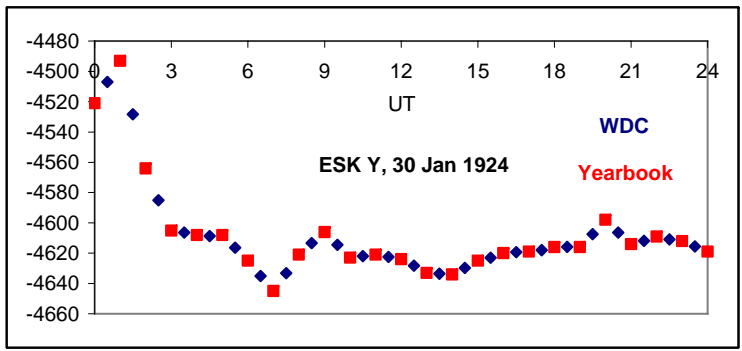

**Figure 25.** The variation of geomagnetic components X, Y, and Z on 30 January 1924 for ESK plotted using the hourly values supplied by the WDCs (blue diamonds) and given in the original observatory yearbook (red squares). It is unmistakable that the WDC data is simply interpolated between the whole hourly data given in the yearbook.

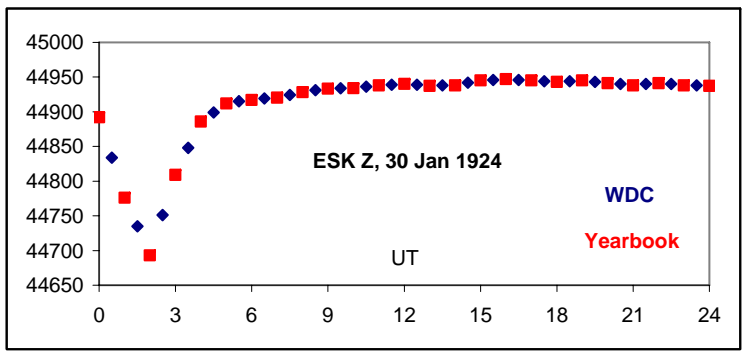

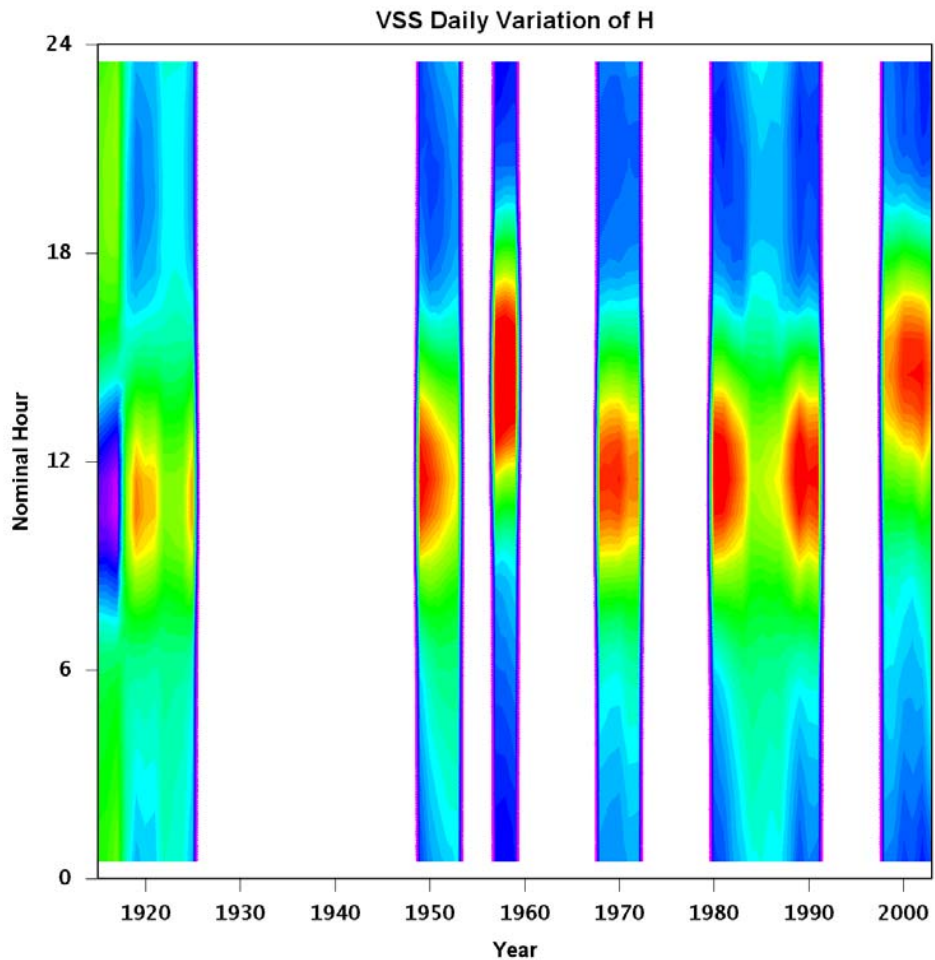

**Figure 26.** The diurnal variation of the horizontal component through the years for Vassouras (VSS) near Rio de Janeiro. The observatory has been in continuous operation since 1915 and is important as the longest running station in its longitude sector in the Southern Hemisphere. The plot is a contour-plot of the variation of the H-component about its daily mean as a function of the hour as given in the WDC-data (the "nominal" hour). Colors from purple/blue to orange/red signify the range from low (negative) to high (positive) values. White areas show where data is missing from the WDC archive.

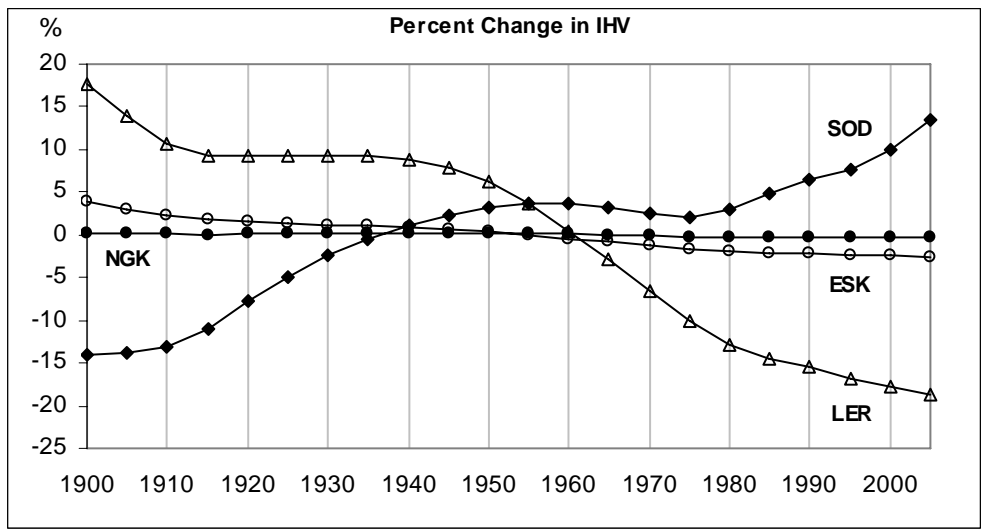

**Figure 27.** (Upper) Percentage change relative to the mean values over 1900-2005 of *IHV* expected for LER, SOD, ESK, and NGK [using eq.(3)] resulting from actual changes in corrected geomagnetic latitude for these stations..

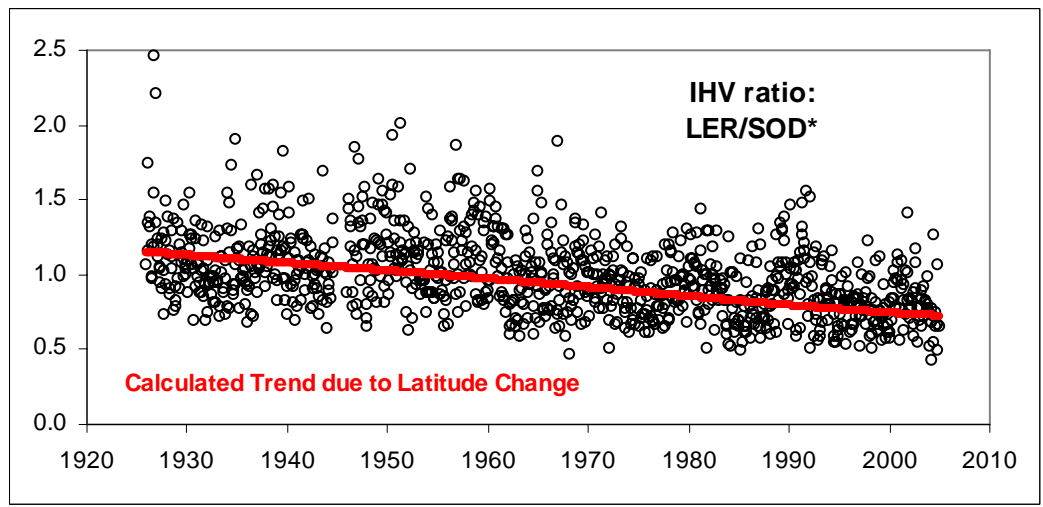

(Lower) The ratio LER/(scaled SOD) for each Bartels rotation since 1926 of calculated *IHV* from the actual data for LER and SOD (the latter scaled by 0.2579 to match the mean of LER). The red line shows the ratio expected (from eq.(3)) due solely to the changing latitudes.

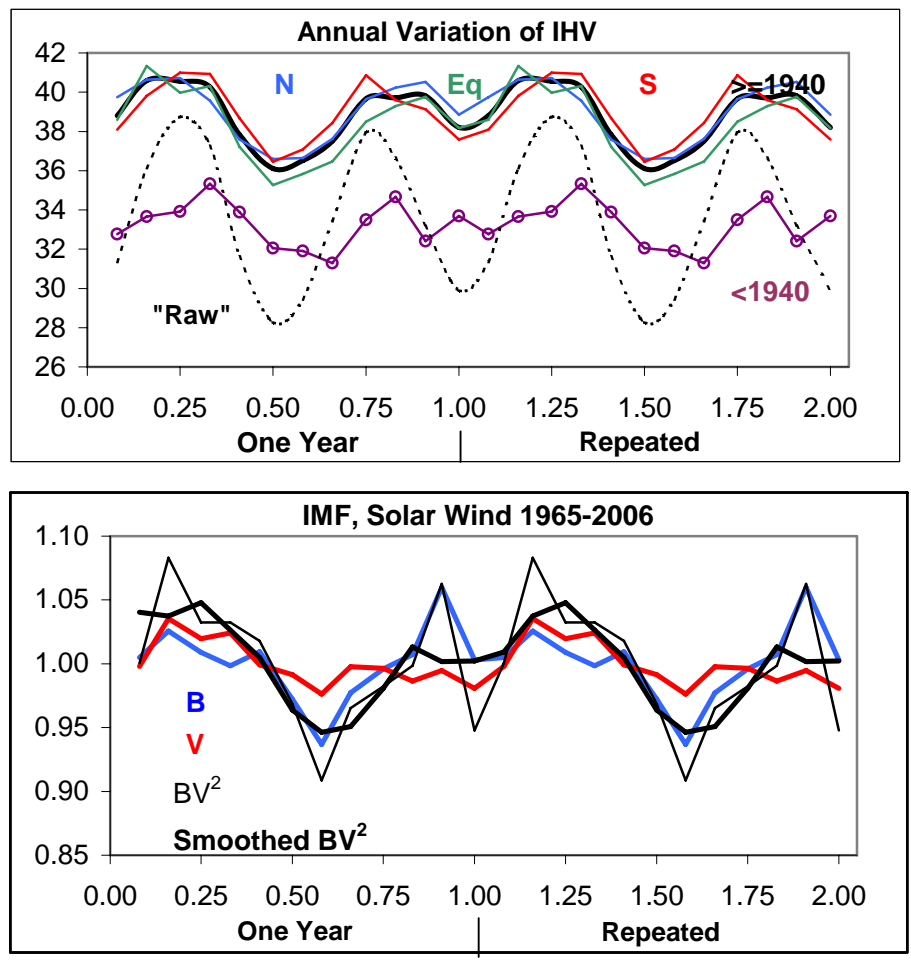

**Figure 28.** (Upper) Annual variation of *IHV* for the composite Northern Hemisphere (blue), Southern Hemisphere (red), and Equatorial (green) series for years 1940 to the present. The average of these three series is shown with a thick black curve. Below this curve we show (purple curve with open circles) the average annual variation of the full *IHV* series for years before 1940 where the data is sparser, especially for the Southern Hemisphere. The dotted curve shows the variation of the "raw" *IHV* (*i.e.* not corrected for the dipole tilt). To better show the annual variation we have repeated the curves for yet another year in the right-hand portion of the Figure. (Lower) Average annual variation of IMF *B* (blue), solar wind speed *V* (red), and the product $BV^2$ (thin black) relative to their mean values for 1965-2006. The heavy black curve shows a three-point running mean of normalized $BV^2$. It would seem that most of the annual variation of *IHV* can be explained simply as variation of the driving $BV^2$.